\documentclass[11pt]{article}
\usepackage{amsmath,amssymb,color,graphics,epsfig,cite,amsmath,mathrsfs}

\usepackage{amsfonts}

\newcommand{\be}{\begin{equation}}
\newcommand{\ee}{\end{equation}}
\newcommand{\bea}{\setlength\arraycolsep{2pt} \begin{eqnarray}}
\newcommand{\eea}{\end{eqnarray}}
\newcommand{\nn}{\nonumber}

\def\fft#1#2{{\frac{#1}{#2}}}

\def\0{{\sst{(0)}}}
\def\1{{\sst{(1)}}}
\def\2{{\sst{(2)}}}
\def\3{{\sst{(3)}}}
\def\4{{\sst{(4)}}}
\def\5{{\sst{(5)}}}
\def\6{{\sst{(6)}}}
\def\7{{\sst{(7)}}}
\def\8{{\sst{(8)}}}
\def\sst#1{{\scriptscriptstyle #1}}
\def\oneone{\rlap 1\mkern4mu{\rm l}}

\begin{document}

\begin{center}
{\Large {\bf General Eight-charge Dyonic Rotating AdS Black Holes \\ in Gauged STU Supergravity}}

\vspace{20pt}

Liang Ma and H. L\"u

\vspace{10pt}

{\it Center for Joint Quantum Studies, Department of Physics,\\
School of Science, Tianjin University, Tianjin 300350, China }

\vspace{40pt}

\underline{ABSTRACT}
\end{center}

We present the most general exact solution for non-extremal rotating AdS black holes in four-dimensional gauged STU supergravity, with four independent electric and four independent magnetic charges. We identify various consistent truncations of the STU model, both supersymmetric or non-supersymmetric, and construct the corresponding solutions. We compare our results with the known special cases in the literature. We analyse the global structure, derive the black hole thermodynamics, evaluate the Euclidean on-shell action, derive the entropy-product relations and analyse the BPS-like limits. Two features deserve emphasis: (1) The general eight-charge rotating AdS black hole exhibits a conical singularity associated with the azimuthal angle; (2) Nontrivial scalar contributions enter the black hole first law. The latter feature has also been noted previously.

\vfill {\footnotesize maliang0@tju.edu.cn\ \ \ mrhonglu@gmail.com}


\thispagestyle{empty}
\pagebreak

\tableofcontents

\newpage

\section{Introduction}
\label{sec:intro}

One salient feature of string theory is its rich structure of duality symmetries, many of which have no conventional quantum field theory (QFT) counterparts. The early-discovered T-duality \cite{Buscher:1987sk}, which relates strings propagating on circles of large and small radii in the target space, is hard to realize in a conventional QFT. Sen's S-duality \cite{Sen:1994fa}, a strong-weak duality of strings, provides a new framework for studying string theory non-perturbatively. The promotion of the Cremmer-Julia global symmetry groups \cite{Cremmer:1978ds,Julia:1980gr} of supergravities to U-duality \cite{Hull:1994ys} ultimately unifies all superstring theories in the framework of M-theory \cite{Witten:1995ex}, where eleven-dimensional supergravity \cite{Cremmer:1978km} and supermembrane action \cite{Bergshoeff:1987cm} play indispensable roles.

The ${\cal N}=2$, $D=4$ STU supergravity model \cite{Duff:1995sm} is a consistent truncation of M-theory and it captures all the three types of string dualities. It can be obtained from a Kaluza-Klein (KK) reduction on a two-torus $T^2$ of the $D=6$ minimal supergravity coupled to one tensor multiplet. The $SO(2,2)$ global symmetry is enhanced to $SL(2,\mathbb R)^3$ because that the string theory three-form field strength is Hodge dual to an axion in four dimensions. This axion and the string dilaton, form an additional $SL(2,\mathbb R)/SO(2)$ complex scalar. The rich $SL(2,\mathbb R)^3$ triality symmetry makes the STU supergravity model a particularly simple but nontrivial setting in which to probe the effects of these three important string dualities.

The bosonic sector of STU supergravity consists of four $U(1)$ gauge fields and three complex scalars.
The theory is particularly useful for studying the properties of charged black holes in string theory, both non-extremal and extremal (in the BPS limit). For suitable charge configurations, the axions can be decoupled and the theory reduces to special class of generalized Einstein-Maxwell-dilaton (EMD) theories. (We refer to a theory with one dilaton and one gauge field as EMD, whereas a theory with multiple dilatons and gauge fields is referred to as generalized EMD.) The KK theory is a special EMD theory and its static dyonic black hole, known as the Gibbons-Wiltshire black hole, was constructed in the early days of string theory \cite{Gibbons:1985ac}. It turns out that the equations can be recast as one-dimensional Toda equations associated with group $SL(3,\mathbb R)$. This provides the first example of the connection between black hole and $p$-brane solutions, on the one hand, and Liouville and Toda equations, on the other, a connection that has been extensively studied \cite{Lu:1996hh,Lu:1996jr,Lu:1996jk,Lu:1997rd, Goldstein:2005hq,Lu:2013uia,Ivashchuk:2002ge, Ivashchuk:2013jja,Zadora:2017qeo,Lu:2026lkv}.

Purely electric or magnetic black holes in EMD theories, governed by the relatively simple Liouville equations, were also constructed in the early days, known as the GMGHS black holes \cite{Gibbons:1987ps,Garfinkle:1990qj}. They serve as prototypes for the more general string or $p$-brane solitons in string theory \cite{Duff:1994an}. These black holes can be lifted to higher dimensions to become $p$-branes, such as strings \cite{Dabholkar:1990yf}, M-branes \cite{Gueven:1992hh} or D3-branes \cite{Duff:1991pea}. The multi-charge black holes of a suitable generalized EMD theory may have their origins in intersecting $p$-branes in higher dimensions \cite{Papadopoulos:1996uq,Klebanov:1996mh}. Classifications of $p$-branes and black holes can be found in \cite{Lu:1995cs,Behrndt:1996pm,Lu:1997hb}. Remarkably, it has recently been shown that the thermodynamics of these black holes or $p$-branes can be derived without knowing their solutions at all \cite{Yang:2025rud,Lu:2025eub,Han:2026bim, Wu:2026udi,Huang:2026ivj}.

Supergravities in diverse dimensions \cite{Salam:1989ihk} can have large manifest global symmetries such as the Cremmer-Julia groups. A solution obtained from another known solution by applying such a manifest global symmetry is typically not regarded as a new solution, but rather as a point that lies in the orbit of the global symmetry group. On the other hand, these global symmetries can be viewed as hidden symmetries from the higher-dimensional point of view, and uplifting such solutions in the same orbit may generate new higher-dimensional solutions. For example, applying the $SO(2)$ electricmagnetic duality of the Einstein-Maxwell theory, we can obtain the dyonic solution from an electrically charged Reissner-Nordstr\"om (RN) black hole, and hence we shall not regard the dyonic solution as a new independent solution from the original RN black hole. On the other hand, if we perform the KK reduction of the Einstein-Maxwell theory, the resulting three-dimensional theory has an $SU(2,1)$ global symmetry. We can employ this global symmetry, a hidden symmetry from the four-dimensional point of view, to generate the RN black hole from the Schwarzschild metric. In this case, we regard the RN solution as a new solution, compared to the Schwarzschild metric. This is of course precisely the underlying symmetry behind the Ernst-Harrison transformation \cite{Ernst:1967wx,Harrison:1968wue}, which is a powerful solution-generating procedure.

In four dimensions, a $U(1)$ gauge field can carry both electric and magnetic charges, and the resulting black hole is dyonic. However, the term ``dyon'' is sometimes used imprecisely in the literature. For example, in generalized EMD theories with multiple $U(1)$ gauge fields, one can construct charged black holes charges are carried by different gauge fields. In the literature, black holes with electric and magnetic charges carried by different gauge fields are also referred to as dyonic. In this paper, we shall refer to these black holes as being ``purely electric'' for simplicity. This is because if a gauge field carries only electric or magnetic charges, we can always Hodge dualize this field so that the dualized gauge field is electric. Consequently, there always exists a Hodge dual version of the theory so that the black hole is indeed purely electric.  We refer to a solution as dyonic only when at least one gauge field simultaneously carries both electric and magnetic charges.

To clarify the above issues further, we consider a concrete example: $T^2$ reduction of pure gravity in six dimensions. The four-dimensional theory contains two KK gauge fields, ${\cal A}_\1^1$ and ${\cal A}_\1^2$, as well as a scalar coset $\mathbb R\times SL(2,\mathbb R)/SO(2)$. The most general black hole solution contains two electric and two magnetic charges, namely $(q_1,q_2)$ and $(p_1,p_2)$. Such a general four-charge solution is dyonic, but configurations with $q_1=0=p_2$ or $q_2=0=p_1$ are regarded as purely electric in the context of this paper. The $SL(2,\mathbb R)$ can be decomposed into its Borel subgroup and the compact subgroup $SO(2)$. The Borel subgroup acts on the scalar moduli, and the compact $SO(2)$ subgroup rotates the two gauge fields $({\cal A}_\1^1, {\cal A}_\1^2)$. This leads to the concept of the minimal seed from which the most general charge configurations can be generated by the action of the global symmetry group. In this KK theory from the $T^2$ reduction, such a seed should carry at least three charges, e.g.~ $(q_1,q_2,p_1,0)$. The corresponding black hole is then dyonic.

For charged black holes in the STU supergravity model, the most general charge configurations involve four independent electric and four independent magnetic charges. Owing to the $SL(2,\mathbb R)^3$ global symmetry, the maximal seed involves five charges, e.g.~four electric and one magnetic. Such asymptotically flat five-charge dyonic static black holes were constructed in \cite{Cvetic:1995kv,Cvetic:1996zq}. Four-charge purely electric rotating black holes were initiated in \cite{Cvetic:1996kv}, and finalized in \cite{Chong:2004na}. Dyonic KK rotating black holes were constructed in \cite{Rasheed:1995zv,Larsen:1999pp}. In \cite{Chow:2013tia, Chow:2014cca}, the most general asymptotically flat, dyonic rotating black hole was constructed, which carries four independent electric charges and four independent magnetic charges, i.e.~eight electromagnetic charges in total. The authors indeed pointed out that a rotating black hole with only five independent electromagnetic charges suffices to generate the most general STU solution through U-duality. This is precisely a continuation of the static example in \cite{Cvetic:1996zq}.

In this paper, we are primarily concerned with charged black holes in gauged STU supergravity, in which a scalar potential for the three complex scalar is introduced via the gauging. The vacuum becomes an anti-de Sitter (AdS) spacetime, with a negative cosmological constant determined by the gauge coupling.
Inspired by the AdS/CFT correspondence \cite{Maldacena:1997re}, there has been an extensive effort to constructing AdS black holes in gauged STU supergravity. The static, purely electric, four-charge AdS black hole was constructed in \cite{Duff:1999gh}. It can be viewed as the near-horizon geometry of a rotating membrane when uplifted to $D=11$ \cite{Cvetic:1999xp}. The  pairwise-equal rotating AdS black hole was constructed in \cite{Chong:2004na}. The single-charge and two-charge rotating AdS black holes were constructed in \cite{Chow:2010sf} and \cite{Chow:2010fw}, respectively. Recently, the four-charge purely electric rotating AdS black holes were constructed \cite{Ma:2026osr,Wu:2026jwa}. The first example of a dyonic AdS black hole was constructed in the AdS KK theory \cite{Lu:2013ura}. The general static four-charge dyonic AdS black hole was presented in \cite{Chow:2013gba}, which also generalized the pairwise-equal rotating AdS black hole of \cite{Chong:2004na} to the dyonic case.

In this paper, we present the most general dyonic rotating AdS black holes in gauged STU supergravity, carrying four independent electric charges and four independent magnetic charges, thereby encompassing all the aforementioned special cases. The paper is organized as follows. In Section \ref{sec:theory}, we present the Lagrangian of gauged STU supergravity and derive the equations of motion. We also classify the various consistent truncations of gauged STU supergravity. This is motivated by the complexity of our eight-charge solution, as well as the fact that solutions of truncated theories can be of interest in their own right.  In Section \ref{sec:8charge}, we present the most general solution. In Section \ref{sec:known}, we consider the pairwise-equal truncations and compare the results with the known solutions in the literature.

In Sections \ref{sec:special1}, \ref{sec:special2}, and \ref{sec:special3}, we consider three different truncations and present the corresponding reduced solutions. In Section \ref{sec:thermo}, we study the global structure, and derive the black hole thermodynamics. In Section \ref{sec:onshell}, we compute the Euclidean on-shell action and derive the corresponding free energy.
In Section \ref{sec:product}, we derive the entropy-product rule and also take the BPS-type limit.
We conclude the paper in Section \ref{sec:con}.

\section{Ungauged and gauged STU supergravities}
\label{sec:theory}

\subsection{The Lagrangian for the bosonic sector}

The ungauged STU supergravity model is an ${\cal N}=2$, $D=4$ supergravity coupled to three vector multiplets. The bosonic sector consists of the metric, four $U(1)$ gauge fields and three neutral complex scalars $\tau_i = \chi_i + {\rm i} e^{-\varphi_i}$, with $i=1,2,3$. The theory is invariant under three independent global $SL(2,\mathbb R)$ transformations, associated with the S-, T- and U-dualities of string theory \cite{Duff:1995sm}. The theory can also be gauged, in which case the fermions are charged under the $U(1)^4$ gauge fields. The bosonic Lagrangian takes the form
\be
{\cal L}_4= (R - V)\, {*\oneone} -\frac{1}{2}\sum_{i=1}^3 ( {*d}\varphi_i\wedge d\varphi_i+e^{2\varphi_i} {*d}\chi_i\wedge d\chi_i)+ {\cal L}_{\rm K}\,,
\ee
where we give the explicit kinetic terms for the three complex scalars. $V$ is the scalar potential for the three complex fields, generated by the gauging:
\be
V=-g^2\sum_{i=1}^3\big(2\cosh\varphi_i+\chi_i^2e^{\varphi_i}\big)\,.\label{genV}
\ee
Here, $g$ is the gauge coupling of the fermions. The scalar potential gives rise to an AdS vacuum with a cosmological constant $\Lambda=-3g^2$.

The term ${\cal L}_{\rm K}$ denotes the kinetic terms of the four $U(1)$ fields, which have non-trivial couplings with the three complex scalars. The Lagrangian term is independent of the gauge coupling $g$, as all the gauge fields are Abelian. Thus, ${\cal L}_{\rm K}$ is the same as the term in the corresponding ungauged supergravity. However, there are a few subtleties involving this statement. For ungauged supergravity with $g=0$, in which all the fields are neutral, we can Hodge dualize any of the four $U(1)$ fields. Not only are the resulting bosonic equations of motion equivalent, but the supersymmetric transformation rules are also equivalent since only the field strengths of the $U(1)$ fields enter the supersymmetry transformation rules. When the theory is gauged, the bosonic sector is still neutral, but the fermions are charged, with non-minimal couplings to the gauge potentials. The supersymmetry transformation rules also contain gauge potentials. Thus, there is a unique formulation for the ${\cal L}_{\rm K}$ and we cannot Hodge dualize the whole theory. One way to determine the explicit expression of ${\cal L}_{\rm K}$, besides a direct supergravity construction, is to perform a $U(1)^4$ truncation of the maximal ${\cal N}=8$, $D=4$ gauged supergravity, whose local gauge group is \(SO(8)\).

On the other hand, when constructing classical solutions where the fermions are all turned off, the contributions to the equations of motion from ${\cal L}_{\rm K}$ are equivalent under Hodge dualizations of the gauge fields, even for the gauged theory with $g\ne 0$. It turns out that expression for the ${\cal L}_{\rm K}$ from the $U(1)^4$ truncation is the most complicated one. It is thus advantageous to choose a simpler form of the Lagrangian \(\mathcal{L}_{\rm K}\) when deriving the equations of motion via some appropriate Hodge dualizations. Two simpler choices are available. One is to write ${\cal L}_{\rm K}$ in the form obtained from the $T^2$ reduction of the six-dimensional bosonic string. The other is to derive ${\cal L}_{\rm K}$ from the KK circle reduction of the five-dimensional STU model. Next, we shall give ${\cal L}_{\rm K}$ in all these cases.

\bigskip
\noindent{\bf Case 1: $T^2$ reduction of $D=6$ string}
\medskip

The bosonic sector of the ungauged STU supergravity can be obtained simply through a $T^2$ reduction of the $D=6$ bosonic string, retaining a pair of winding modes $(A_{\1,1},A_{\1,2})$ and a pair of KK modes $(\mathcal{A}_{\1}^1,\mathcal{A}_{\1}^2)$. This is because the bosonic sector of $D=6$ minimal supergravity coupled to a tensor multiplet is exactly the same as the bosonic string.
Following the convention of \cite{Chong:2004na}, the relevant ${\cal L}_{\rm K}$ is given by
\bea
\mathcal{L}_{\rm K} &=&-\frac{1}{2}e^{-\varphi_1}\Big(
e^{\varphi_2-\varphi_3} {*F}_{\2,1}\wedge F_{\2,1}+e^{\varphi_2+\varphi_3} {*F}_{\2,2}\wedge F_{\2,2}+e^{-\varphi_2+\varphi_3} {*\mathcal F}_{\2}^1\wedge \mathcal{F}_{\2}^1
\cr
&&+e^{-\varphi_2-\varphi_3} {*\mathcal F}_{\2}^2\wedge \mathcal{F}_{\2}^2
\Big)-\chi_1\big(F_{\2,1}\wedge \mathcal{F}_{\2}^1+F_{\2,2}\wedge \mathcal{F}_{\2}^2\big)\,,\label{lag}
\eea
where the field strengths and gauge potentials are related by
\bea
F_{\2,1}&=&dA_{\1,1}-\chi_2d\mathcal{A}_{\1}^2\,,\qquad
F_{\2,2}=dA_{\1,2}+\chi_2d\mathcal{A}_{\1}^1-\chi_3dA_{\1,1}+
\chi_2\chi_3d\mathcal{A}_{\1}^2\,,\cr
\mathcal{F}_{\2}^1&=&d\mathcal{A}_{\1}^1+\chi_3d\mathcal{A}_{\1}^2\,,\qquad
\mathcal{F}_{\2}^2 = d\mathcal{A}_{\1}^2\,.\label{lagFA}
\eea
This Lagrangian was initially studied in \cite{Duff:1995sm} and it has the advantage that it makes clear how the four $U(1)$ fields and their Hodge duals form respective pairwise doublets under the three $SL(2,\mathbb R)$ symmetry transformations. (See \cite{Duff:1995sm} for details.)

\bigskip
\noindent{\bf Case 2: $S^1$ reduction of $D=5$ STU supergravity}
\medskip

The five-dimensional STU supergravity is minimal supergravity coupled to two vector multiplets. It can itself be obtained by first reducing the \(D=6\) string on a circle and then Hodge-dualizing the three-form field strength.. The theory thus contains three gauge fields. Upon circle reduction, they, together with the KK vector, yield the four $U(1)$ fields in the $D=4$ STU supergravity. This reduction route amounts to decomposing the $T^2$ reduction into two successive $S^1$ reductions, with a Hodge dualization of the 3-form field strength between them. The new feature is that the three gauge fields in $D=5$ are totally symmetric, equivalent to each other. The four-dimensional theory thus has a natural 3+1 splitting of the four gauge fields, in which the three gauge fields can be set equal upon a further consistent truncation. This theory can also be obtained directly from \eqref{lag} by Hodge dualization. We perform an electromagnetic duality transformation on \(A_{\1,2}\) in \eqref{lag} and obtain \(\tilde A_{\1,2}\) via the duality relation:
\be
d\tilde A_{\1,2}-\chi_1\mathcal F_{\2}^{2}=e^{-\varphi_1+\varphi_2+\varphi_3}\, {*F}_{\2,2}\,,
\ee
which we can use to solve for $F_{\2,2}$. At this point, the Lagrangian \eqref{lag} can be rewritten as
\bea
{\mathcal L}_{\rm K} &=&
-\frac12e^{-\varphi_1-\varphi_2-\varphi_3}
\big(
e^{2\varphi_1} {*\tilde F}_{\2}^2\wedge\tilde F_{\2}^2
+e^{2\varphi_2} {*F}_{\2,1}\wedge F_{\2,1}
+e^{2\varphi_3} {*\mathcal F}_{\2}^1\wedge\mathcal F_{\2}^1\cr
&&+ {*\mathcal F}_{\2}^2\wedge\mathcal F_{\2}^2
\big)+\big(\chi_3F_{\2,1}-\chi_2\mathcal F_{\2}^1
+\chi_2\chi_3\mathcal F_{\2}^2\big)
\wedge(\tilde F_{\2}^2+\chi_1\mathcal F_{\2}^2)-\chi_1F_{\2,1}\wedge\mathcal F_{\2}^1\,,\nn\\
\tilde F_{\2}^2&=&d\tilde A_{\1,2}-\chi_1\mathcal F_{\2}^2\,,\qquad
 F_{\2,1}=dA_{\1,1}-\chi_2\mathcal F_{\2}^2\,,\cr
\mathcal F_{\2}^1&=&d\mathcal A_{\1}^1+\chi_3\mathcal F_{\2}^2\,,\qquad
 \mathcal F_{\2}^2=d\mathcal A_{\1}^2\,.\label{lag2}
\eea
In this formulation, the field strength $F_{\2}^2$ is singled out in the Lagrangian, whilst the remaining three field strengths are written in the symmetric form. It is worth noting that the $D=5$ STU supergravity can also be gauged, and considerable effort has been devoted to the construction of charged rotating AdS black holes \cite{Cvetic:1996xz,Cvetic:2004hs,Chong:2005da, Chong:2005hr,Chong:2006zx,Wu:2011gq}.

\bigskip
\noindent{\bf Case 3: $U(1)^4$ truncation of $SO(8)$}
\medskip

As discussed above, STU gauged supergravity can be obtained as a consistent $U(1)^4$ Cartan-group truncation of the maximal gauged supergravity in four dimensions, whose local gauge group is \(SO(8)\). Its Lagrangian term ${\cal L}_{\rm K}$ is related to the one given from $T^2$ reduction of the bosonic string through the Hodge dualization of two $U(1)$ fields. As shown in \cite{Cvetic:1999xp}, by introducing the Lagrange multiplier terms \(dA_{\1,1}\wedge d\tilde A_{\1,1}\) and \(d\mathcal A_{\1}^1\wedge d\tilde{\mathcal A}_{\1}^1\), we can implement the electromagnetic duality
\bea
d \tilde{A}_{\1,1}
&=&e^{-\varphi_1+\varphi_2-\varphi_3} {*(dA_{\1,1}-\chi_2\,d\mathcal A_{\1}^2)}
-\chi_3\mathcal{G}_{\2}+\chi_1\,d\mathcal A_{\1}^1\,,\cr
d\tilde{\mathcal A}_{\1}^1
&=&e^{-\varphi_1-\varphi_2+\varphi_3} {*(d\mathcal A_{\1}^1+\chi_3\,d\mathcal A_{\1}^2)}
+\chi_2 \mathcal{G}_{\2}+\chi_1\,dA_{\1,1}\,,
\eea
where
\be
\mathcal{G}_{\2}
=e^{-\varphi_1+\varphi_2+\varphi_3} {*( dA_{\1,2}+\chi_2d\mathcal A_{\1}^1-\chi_3dA_{\1,1} +\chi_2\chi_3d\mathcal A_{\1}^2)}\,.
\ee
Upon performing the electromagnetic duality transformation on the two $U(1)$ fields, the Lagrangian \eqref{lag} can be recast into the form of the \(U(1)^4\) truncation of \(SO(8)\).  Specifically, we solve for
$(dA_{\1,1}, d\mathcal A_{\1}^1)$ in terms of $(d\tilde A_{\1,1}, d\tilde{\mathcal A}_{\1}^1)$. The resulting four field strengths are then given by
\bea
\mathcal F_{\2}^2&=&d\mathcal A_{\1}^2\,,\qquad
 \tilde{F}_{\2,1}
=d \tilde{A}_{\1,1}+\chi_1\chi_3\mathcal F_{\2}^2\,,\qquad
\tilde{\mathcal F}_{\2}^1
=d\tilde{\mathcal A}_{\1}^1-\chi_1\chi_2\mathcal F_{\2}^2,\nn\\
F_{\2,2}
&=&dA_{\1,2}-\chi_2\chi_3\mathcal F_{(2)}^2+\frac{e^{\varphi_1}}{1+\chi_1^2e^{2\varphi_1}}
{*(\chi_3e^{-\varphi_2+\varphi_3}\tilde F_{\2,1}
-\chi_2e^{\varphi_2-\varphi_3}\tilde{\mathcal F}_{\2}^1)}\nn\\
&&+\frac{\chi_1e^{2\varphi_1}}{1+\chi_1^2e^{2\varphi_1}}
(\chi_2\tilde F_{\2,1}
-\chi_3\tilde{\mathcal F}_{\2}^1)\,.
\eea
Since the kinetic term of \(F_{\2,2}\) is very complicated, we introduce a new notation \(\bar{F}_{\2,2}\), defined by
\bea
\bar{F}_{\2,2}=(1+\chi_1^2e^{2\varphi_1})\frac{(1+\sum_{i=1}^3\chi_i^2e^{2\varphi_i})\,
{*F}_{\2,2}-2\prod_{i=1}^3\chi_ie^{\varphi_i}F_{\2,2}}{(1+\sum_{i=1}^3
\chi_i^2e^{2\varphi_i})^2+4\prod_{i=1}^3\chi_i^2e^{2\varphi_i}}\,.
\eea
The final Lagrangian becomes
\bea
{\mathcal L}_{\rm K}&=&-\frac{e^{\varphi_1}}{2(1+\chi_1^2e^{2\varphi_1})}
\big(e^{-\varphi_2+\varphi_3}*\tilde F_{\2,1}\wedge\tilde F_{\2,1}
+e^{\varphi_2-\varphi_3}*\tilde{\mathcal F}_{\2}^1 \wedge\tilde{\mathcal F}_{\2}^1
\big)\cr
&&-\frac{e^{-\varphi_1}}{2}(e^{-\varphi_2-\varphi_3}
*\mathcal F_{\2}^2\wedge\mathcal F_{\2}^2
+ e^{\varphi_2+\varphi_3}F_{\2,2}\wedge\bar{F}_{\2,2})
+\frac{\chi_1e^{2\varphi_1}}{1+\chi_1^2e^{2\varphi_1}}
\tilde F_{\2,1}\wedge\tilde{\mathcal F}_{\2}^1\cr
&&
+\mathcal F_{\2}^2\wedge \Big[ \chi_2d\tilde A_{\1,1}
-\chi_3d\tilde{\mathcal A}_{\1}^1
-\chi_1\big(dA_{\1,2}-\chi_2\chi_3\mathcal F_{\2}^2\big)
\Big]\,.\label{gauged STU dual}
\eea
It is now clear that the actual Lagrangian for the gauged STU supergravity model, namely \eqref{gauged STU dual}, is much more complicated compared than the other formulations. This Lagrangian is not particularly convenient for deriving the equations of motion when simpler alternatives are available.  Therefore, it is common practice to use the equivalent Lagrangians \eqref{lag} or \eqref{lag2} to derive the equations for constructing rotating black hole solutions. In fact, we are not aware of the $U(1)^4$ Lagrangian \eqref{gauged STU dual} having appeared in the literature.

\bigskip
{\noindent\bf Summary}
\medskip

We have so far presented three Lagrangians and explained their origins. These Lagrangians are equivalent at the purely bosonic level, since they are related to each other by some appropriate Hodge dualization. In this paper, we use the Lagrangian obtained from the $T^2$ reduction of the bosonic string as our starting point, and obtain the Lagrangians associated with the $S^1$ reduction of $D=5$ STU theory and the $U(1)^4$ truncation of the $SO(8)$ gauged supergravity through suitable Hodge dualizations.
We introduce new notation of gauge fields after Hodge dualization, by putting a tilde symbol on top of the original ones. Here we summarize the notations as follows
\bea
T^2 \rightarrow S^1:&& \{A_{\1,1}, A_{\1,2}, {\cal A}_\1^1, {\cal A}_\1^2\} \rightarrow  \{A_{\1,1}, \tilde A_{\1,2}, {\cal A}_\1^1, {\cal A}_\1^2\}\,,\cr
T^2 \rightarrow U(1)^4:&& \{A_{\1,1}, A_{\1,2}, {\cal A}_\1^1, {\cal A}_\1^2\} \rightarrow  \{\tilde A_{\1,1}, A_{\1,2}, \tilde {\cal A}_\1^1, {\cal A}_\1^2\}\,.
\eea
Specifically, we use the $T^2$ reduction theory as our standard notation for the gauge fields, and a gauge field is denoted with a tilde one after Hodge dualization. Thus the $S^1$ reduction theory is the one obtained from performing Hodge dualization on $A_{\1,2}$ to become $\tilde A_{\1,2}$. The $U(1)^4$ truncation theory is the one obtained from Hodge dualization on two gauged fields $(A_{\1,1},{\cal A}_\1^1)$ to become $(\tilde A_{\1,1},\tilde {\cal A}_\1^1)$. One consequence is that a solution in the $U(1)^4$ theory with four electric charges can equivalently be described as carrying two electric and magnetic charges in the $T^2$ reduction theory, and one electric and three magnetic charges in the $S^1$ reduction theory. As explained already in the introduction, a black hole solution with each gauge field carrying only one type of charge, either electric or magnetic, is referred to as a purely electric black hole.

To compare the defining characteristics of the three different forms of ${\cal L}_{\rm K}$, we simplify the Lagrangians as much as possible, setting all the axions to zero, so that each Lagrangian reduces to that of a generalized EMD theory, with four gauge fields and three dilatons. The Lagrangians all take the form
\be
{\cal L}_{\rm K} \sim -\fft12 \sum_{i=1}^4 e^{\vec a_i \cdot\vec\varphi}\, {*F_{\2,i}}\wedge F_{\2,i}\,.
\label{genEMD}
\ee
For the three different representations, the dilaton vectors are
\bea
U(1)^4:&& \vec a_1=(1,-1,1),\quad \vec a_2=(-1,1,1)\,\quad \vec a_3 = (1,1,-1),\quad \vec a_4=(-1,-1,-1)\,,\cr
T^2:&& \vec a_1=(-1,1,-1),\quad \vec a_2=(-1,1,1)\,\quad \vec a_3 = (-1,-1,1),\quad \vec a_4=(-1,-1,-1)\,,\cr
S^1:&& \vec a_1=(-1,1,-1),\quad \vec a_2=(1,-1,-1)\,\quad \vec a_3 = (-1,-1,1),\quad \vec a_4=(-1,-1,-1)\,.
\eea
The Static AdS black hole carrying four electric charges was constructed in \cite{Duff:1999gh} in the $U(1)^4$ theory. Its uplift to $D=11$ describes the near-horizon geometry of rotating membranes, with the four electric charges corresponding to four angular momenta of the four orthogonal rotations in $S^7$ \cite{Cvetic:1999xp}. The four electric charges become two electric and two magnetic charges in the $T^2$ reduction theory and one electric and three magnetic charges in the $S^1$ reduction theory.

\subsection{Residual $SO(2)^3$ global symmetry}

When the gauge coupling $g$ vanishes, the ungauged theory is invariant under three global $SL(2,\mathbb R)$ transformations. Its transitive action on the three complex scalars is
\be
\tau_i \rightarrow \tau_i' = \frac{a_i \tau_i + b_i}{c_i \tau_i + d_i}\,,\qquad
\hbox{with}\qquad a_i d_i - b_i c_i=1\,,\qquad i=1,2,3\,.\label{sl2r}
\ee
The four $U(1)$ gauge fields and their Hodge duals form appropriate doublets under these transformations. In the gauged theory, since ${\cal L}_{\rm K}$ is independent of the gauge coupling, the breaking of the global symmetry in the bosonic sector is solely due to the scalar potential $V$ in \eqref{genV}. It can be easily verified that the scalar potential is invariant under the compact subgroups $SO(2)$ subgroups of the three \(SL(2,\mathbb R)\) factors. Consequently, the full bosonic Lagrangian is invariant under the transformation \eqref{sl2r} with
\be
\Lambda_i=
\left(
  \begin{array}{cc}
    a_i & b_i \\
    c_i & d_i \\
  \end{array}
\right)=
\left(
  \begin{array}{cc}
    \cos\theta_i & -\sin\theta_i \\
    \sin\theta_i & \cos\theta_i \\
  \end{array}
\right)\,,\qquad i=1,2,3\,.
\ee
This implies that the minimal seed required to generate all eight-charge dyonic rotating AdS black holes must carry \(8-3=5\) charges, the same number as in the corresponding ungauged supergravity.

\subsection{Covariant equations of motion}

As was discussed, in the pure bosonic sector, there is more than one way to write the Lagrangian ${\cal L}_{\rm K}$; however, it is to us the one given in \eqref{lag} has the simplest form. We shall derive the equations of motion from the variation based on \eqref{lag}. For convenience, we no longer use differential-form notation here, and we also move the index \(i=1,2\) on the two KK gauge fields \((\mathcal{A}_{\1}^1,\mathcal{A}_{\1}^2)\) from the superscript to the subscript. Furthermore, we denote the closed 2-form field strengths $F_{\2,i}^0=dA_{\1,i}$, $\mathcal{F}_{\2}^{i,0}= d\mathcal{A}_{\1}^i$ by a superscript \(0\) and introduce the following shorthand notation
\be
F_{i,\mu\nu}^2=F_{i,\mu\rho}F_{i,\nu}^{{\color{white}{i,\nu}}\rho}\,.
\ee
The complete equations of motion are given by the vanishing of each of the following quantities:
\bea
E_{\mu\nu}&=&-\frac12g_{\mu\nu}\mathcal L_{4,p}+R_{\mu\nu}
-\frac12\sum_{i=1}^3(\nabla_\mu\varphi_i)(\nabla_\nu\varphi_i)
-\frac12\sum_{i=1}^3e^{2\varphi_i}(\nabla_\mu\chi_i)(\nabla_\nu\chi_i)\cr
&&-\frac12e^{-\varphi_1}\big[
 e^{\varphi_2-\varphi_3}F_{1,\mu\nu}^2
 +e^{\varphi_2+\varphi_3}F_{2,\mu\nu}^2
 +e^{-\varphi_2+\varphi_3}\mathcal F_{1,\mu\nu}^2
 +e^{-\varphi_2-\varphi_3}\mathcal F_{2,\mu\nu}^2\big]\,,\cr
S^\nu_{A_1}&=&\nabla_\mu\big(
 e^{-\varphi_1+\varphi_2-\varphi_3}F_1^{\mu\nu}
 -e^{-\varphi_1+\varphi_2+\varphi_3}\chi_3F_2^{\mu\nu}
 -\frac{1}{2}\chi_1\epsilon^{\mu\nu\rho\sigma}\mathcal F^0_{1,\rho\sigma}\big)\,,\cr
S^\nu_{A_2}&=&\nabla_\mu\big(
 e^{-\varphi_1+\varphi_2+\varphi_3}F_2^{\mu\nu}
 -\frac{1}{2}\chi_1\epsilon^{\mu\nu\rho\sigma}\mathcal F^0_{2,\rho\sigma}\big)\,,\cr
S^\nu_{\mathcal A_1}&=&\nabla_\mu\big(
 e^{-\varphi_1+\varphi_2+\varphi_3}\chi_2F_2^{\mu\nu}
 +e^{-\varphi_1-\varphi_2+\varphi_3}\mathcal F_1^{\mu\nu}
 -\frac{1}{2}\chi_1\epsilon^{\mu\nu\rho\sigma}F^0_{1,\rho\sigma}\big)\,,\cr
S^\nu_{\mathcal A_2}&=&\nabla_\mu\big(
 -e^{-\varphi_1+\varphi_2-\varphi_3}\chi_2F_1^{\mu\nu}
 +e^{-\varphi_1+\varphi_2+\varphi_3}\chi_2\chi_3F_2^{\mu\nu}\cr
&&
 +e^{-\varphi_1-\varphi_2+\varphi_3}\chi_3\mathcal F_1^{\mu\nu}
 +e^{-\varphi_1-\varphi_2-\varphi_3}\mathcal F_2^{\mu\nu}
 -\frac{1}{2}\chi_1\epsilon^{\mu\nu\rho\sigma}F^0_{2,\rho\sigma}\big)\,,\cr
E_{\varphi_1}&=&\Box\varphi_1-e^{2\varphi_1}(\nabla\chi_1)^2
+\frac14e^{-\varphi_1}\big[
 e^{\varphi_2-\varphi_3}F_1^2+e^{\varphi_2+\varphi_3}F_2^2
 +e^{-\varphi_2+\varphi_3}\mathcal F_1^2
 +e^{-\varphi_2-\varphi_3}\mathcal F_2^2\big]-\frac{\partial V}{\partial\varphi_1}\,,\cr
E_{\varphi_2}&=&\Box\varphi_2-e^{2\varphi_2}(\nabla\chi_2)^2
-\frac14e^{-\varphi_1}\big[
 e^{\varphi_2-\varphi_3}F_1^2+e^{\varphi_2+\varphi_3}F_2^2
 -e^{-\varphi_2+\varphi_3}\mathcal F_1^2
 -e^{-\varphi_2-\varphi_3}\mathcal F_2^2\big]-\frac{\partial V}{\partial\varphi_2}\,,\cr
E_{\varphi_3}&=&\Box\varphi_3-e^{2\varphi_3}(\nabla\chi_3)^2
+\frac14e^{-\varphi_1}\big[
 e^{\varphi_2-\varphi_3}F_1^2-e^{\varphi_2+\varphi_3}F_2^2
 -e^{-\varphi_2+\varphi_3}\mathcal F_1^2
 +e^{-\varphi_2-\varphi_3}\mathcal F_2^2\big]-\frac{\partial V}{\partial\varphi_3}\,,\cr
E_{\chi_1}&=&\nabla_\mu\big(e^{2\varphi_1}\nabla^\mu\chi_1\big)+\frac14\epsilon^{\mu\nu\rho\sigma}
 \big[F^0_{1\mu\nu}\mathcal F^0_{1\rho\sigma}
      +F^0_{2\mu\nu}\mathcal F^0_{2\rho\sigma}\big]-\frac{\partial V}{\partial\chi_1}\,,\cr
E_{\chi_2}&=&\nabla_\mu\big(e^{2\varphi_2}\nabla^\mu\chi_2\big)
+\frac12e^{-\varphi_1+\varphi_2-\varphi_3}
    F_1^{\mu\nu}\mathcal F_{2\mu\nu}
-\frac12e^{-\varphi_1+\varphi_2+\varphi_3}
    F_2^{\mu\nu}\mathcal F_{1\mu\nu}-\frac{\partial V}{\partial\chi_2}\,,\cr
E_{\chi_3}&=&\nabla_\mu\big(e^{2\varphi_3}\nabla^\mu\chi_3\big)
+\frac12e^{-\varphi_1+\varphi_2+\varphi_3}
    F_2^{\mu\nu}F_{1\mu\nu}
-\frac12e^{-\varphi_1-\varphi_2+\varphi_3}
    \mathcal F_1^{\mu\nu}\mathcal F_{2\mu\nu}-\frac{\partial V}{\partial\chi_3}\,.
\eea

\subsection{Various consistant truncations}

Our main goal is to construct the most general rotating AdS black holes in gauged STU supergravity. There are many known special solutions in the literature, which were reviewed in the Introduction. The gauged STU theory admits a variety of consistent truncations. In order to compare our general results with the special cases in the literature, we review these consistent truncations. Furthermore, we can also obtain new solutions in the reduced theories, in which the solutions are nontrivial but significantly simpler. Taking the Lagrangian \eqref{lag} as our starting point, we now describe the commonly used truncation schemes together with the resulting theories. These truncations were thoroughly analyzed in \cite{Chow:2014cca}, but we shall also present some new explicit Lagrangians obtained through suitable Hodge dualizations.

\subsubsection{Pairwise-equal truncation}

In the gauged STU theory, the pairwise-equal case is currently the most extensively studied consistent truncation. This reduction originates from the fact that it is always consistent to set the winding mode and the KK mode equal in a KK reduction of string theory on a circle. In the gauged STU supergravity model, following the truncation, the matter sector of the theory consists of two \(U(1)\) gauge fields and one complex scalar. Specifically, starting from the Lagrangian \eqref{lag}, the truncation amounts to setting
\bea
&&\varphi^{\mathrm{STU}}_2=\varphi^{\mathrm{STU}}_3=0\,,\qquad \chi^{\mathrm{STU}}_2=\chi^{\mathrm{STU}}_3=0\,,\qquad \varphi^{\mathrm{STU}}_1=\varphi\,,\qquad \chi^{\mathrm{STU}}_1=\chi\,,\cr
&&A_{\1,1}^{\mathrm{STU}}=\mathcal{A}_{\1}^{1,\mathrm{STU}}=\frac{1}{\sqrt{2}}A_{\1,1}\,,\qquad A_{\1,2}^{\mathrm{STU}}=\mathcal{A}_{\1}^{2,\mathrm{STU}}=\frac{1}{\sqrt{2}}A_{\1,2}\,.
\eea
The reduced Lagrangian becomes \cite{Chong:2004na}
\bea
\mathcal{L}_4&=&\big(R+g^2(4+2\cosh\varphi+e^\varphi\chi^2)\big)\, {*\oneone}
-\frac{1}{2} {*d}\varphi\wedge d\varphi-\frac{1}{2}e^{2\varphi} {*d}\chi\wedge d\chi\cr
&&-\frac{1}{2}e^{-\varphi}\big(
{*F}_{\2,1}\wedge F_{\2,1}+ {*F}_{\2,2}\wedge F_{\2,2}\big) -\frac{1}{2}\chi\big(F_{\2,1}\wedge F_{\2,1}+F_{\2,2}\wedge F_{\2,2}\big)\,,\cr
F_{\2,1}&=&dA_{\1,1}\,,\qquad F_{\2,2}=dA_{\1,2}\,.
\label{pairwise action}
\eea
If we instead start from the Lagrangian \eqref{gauged STU dual} of the $U(1)^4$ theory, the truncation is to take
\bea
&&\varphi_2^{\mathrm{STU}}=\varphi_3^{\mathrm{STU}}
=\chi_2^{\mathrm{STU}}=\chi_3^{\mathrm{STU}}=0\,,
\qquad
\varphi_1^{\mathrm{STU}}=\varphi\,,
\qquad
\chi_1^{\mathrm{STU}}=\chi\,,\cr
&&\tilde{A}_{\1,1}^{\mathrm{STU}}=\tilde{\mathcal{A}}_{\1}^{1,\mathrm{STU}}=
\frac{1}{\sqrt{2}}\tilde{A}_{\1,1}\,,\qquad A_{\1,2}^{\mathrm{STU}}=\mathcal{A}_{\1}^{2,\mathrm{STU}}=\frac{1}{\sqrt{2}}A_{\1,2}\,.
\eea
The reduced Lagrangian is
\bea
\tilde{\mathcal{L}}_4&=&
R\,{*\oneone}-\frac12 {*d}\varphi\wedge d\varphi
-\frac12e^{2\varphi} {*d}\chi\wedge d\chi-\frac12e^{-\varphi} {*F}_{\2,2}\wedge F_{\2,2}\cr
&&-\frac12\frac{e^\varphi}{1+\chi^2e^{2\varphi}}
{*\tilde F}_{\2,1}\wedge\tilde F_{\2,1}-\frac12\chi F_{\2,2}\wedge F_{\2,2}
+\frac12\frac{\chi e^{2\varphi}}{1+\chi^2e^{2\varphi}}
\tilde F_{\2,1}\wedge\tilde F_{\2,1}\cr
&&+g^2\big(4+2\cosh\varphi+\chi^2e^\varphi\big)\,{*\oneone}
\,,\cr
\tilde{F}_{\2,1}&=&d\tilde{A}_{\1,1}\,,\qquad F_{\2,2}=dA_{\1,2}\,.
\label{pairwise action dual}
\eea
This Lagrangian is related to the earlier one \eqref{pairwise action} by an appropriate Hodge dualization. As one would expect, this Lagrangian is the \(U(1)^2\) truncation of \(D=4\), \(\mathcal{N}=4\), \(SO(4)\) gauged supergravity \cite{Gates:1982ct}, which can be consistently obtained from a suitable \(S^7\) reduction of eleven-dimensional supergravity \cite{Cvetic:1999au}.

\subsubsection{EMDA truncation}

The ungauged Einstein-Maxwell-dilaton-axion theory (EMDA) is the dual of the heterotic string and was initially studied by Sen in the context of S-duality \cite{Sen:1994fa}. The gauged EMDA supergravity can be obtained by further truncating the pairwise-equal theory \eqref{pairwise action} by taking
\be
\varphi^{\mathrm{PW}}=\varphi\,,\qquad
\chi^{\mathrm{PW}}=\chi\,,\qquad
A_{\1,1}^{\mathrm{PW}}=0\,,\qquad A_{\1,2}^{\mathrm{PW}}=A_{\1}\,.
\ee
The resulting theory is the standard gauged EMDA theory
\bea
\mathcal L_4&=&\Big(R  +g^2\bigl(4+2\cosh\varphi+\chi^2e^\varphi\bigr)\Big)\, {*\oneone}
-\frac{1}{2} {*d}\varphi\wedge d\varphi
-\frac{1}{2}e^{2\varphi} {*d}\chi\wedge d\chi \cr
&&-\frac{1}{2}e^{-\varphi} {*F}_{\2}\wedge F_{\2} -\frac{1}{2}\chi F_{\2}\wedge F_{\2}\,,
\qquad F_{\2}=dA_{\1} \,.
\eea

\subsubsection{Three-equal truncation}

From the viewpoint of the ungauged theory, the truncated theory can be obtained from a circle reduction of five-dimensional minimal supergravity, in which the three gauge fields of the five-dimensional STU supergravity model are all set equal. It is thus natural to start from the Lagrangian \eqref{lag2} and set the three $U(1)$ gauge fields equal. Specifically, we take
\bea
&&\varphi_1^{\mathrm{STU}}=\varphi_2^{\mathrm{STU}}=\varphi_3^{\mathrm{STU}}
=-\frac{\varphi}{\sqrt{3}}\,,\qquad \chi_1^{\mathrm{STU}}=\chi_3^{\mathrm{STU}}=\frac{\chi}{\sqrt{3}}\,,\qquad \chi_2^{\mathrm{STU}}=-\frac{\chi}{\sqrt{3}}\,,\cr
&&A_{\1,1}^{\mathrm{STU}}=-\tilde{A}_{\1,2}^{\mathrm{STU}}
=\mathcal{A}_{\1}^{1,\mathrm{STU}}=\frac{A_{\1,1}}{\sqrt{3}}\,,\qquad \mathcal{A}_{\1}^{2,\mathrm{STU}}=A_{\1,2}\,.
\eea
The reduced theory involves two $U(1)$ gauge fields, one dilaton and one axion. The corresponding Lagrangian is
\bea
\mathcal{L}&=&R\, {*\oneone}-\frac{1}{2} {*d}\varphi\wedge d\varphi-\frac{1}{2}e^{-\frac{2}{\sqrt{3}}\varphi} {*d}\chi\wedge d\chi
-\frac{1}{2}e^{\frac{\varphi}{\sqrt{3}}} {*F}_{\2,1}\wedge F_{\2,1}\cr
&&-\frac{1}{2}e^{\sqrt{3}\varphi} {*F}_{\2,2}\wedge F_{\2,2}-\frac{\chi}{\sqrt{3}}\Big(
F_{\2,1}\wedge F_{\2,1}-\chi F_{\2,1}\wedge F_{\2,2}+\frac{\chi^2}{3}F_{\2,2}\wedge F_{\2,2}
\Big)\cr
&&+g^2\Big(6\cosh\frac{\varphi}{\sqrt{3}}+\chi^2e^{-\frac{\varphi}{\sqrt{3}}}\Big)\,{*\oneone}\,,\nn\\
F_{\2,1}&=&dA_{\1,1}+\chi dA_{\1,2}\,,\qquad F_{\2,2}=dA_{\1,2}\,.\label{S3 action}
\eea
If instead, we start from \eqref{gauged STU dual}, and perform the truncation
\bea
\tilde A_{\1,1}^{\mathrm{STU}}
=A_{\1,2}^{\mathrm{STU}}
=\tilde{\mathcal A}_{\1}^{1,\mathrm{STU}}
=\frac{\tilde A_{\1,1}}{\sqrt3},
\qquad
\mathcal A_{\1}^{2,\mathrm{STU}}=A_{\1,2}\,,
\eea
the resulting Lagrangian takes a different but equivalent Hodge dual form
\bea
\tilde{\mathcal L}&=&
R\,{*\oneone}-\frac{1}{2}{*d}\varphi\wedge d\varphi
-\frac{1}{2} e^{-\frac{2\varphi}{\sqrt3}} {*d}\chi\wedge d\chi+g^2(6\cosh\frac{\varphi}{\sqrt3}
+\chi^2 e^{-\frac{\varphi}{\sqrt3}})\, {*\oneone}\cr
&&-\frac{e^{-\frac{\varphi}{\sqrt3}}}
{2(1+\frac{4}{3}\chi^2e^{-\frac{2\varphi}{\sqrt3}})}
{*\tilde F}_{\2,1} \wedge\tilde F_{\2,1}-\frac{1}{2}e^{\sqrt3\varphi} {*F}_{\2,2}\wedge F_{\2,2}\cr
&&+\frac{\chi e^{-\frac{2\varphi}{\sqrt3}}}
{\sqrt3(1+\frac43\chi^2e^{-\frac{2\varphi}{\sqrt3}})}
\tilde F_{\2,1}\wedge\tilde F_{\2,1}-\chi F_{\2,2}\wedge\tilde F_{\2,1}
+\frac{2\chi^3}{3\sqrt3}F_{\2,2}\wedge F_{\2,2}\,,\cr
\tilde F_{\2,1}&=&d\tilde A_{\1,1}+\frac{\chi^2}{\sqrt3}dA_{\1,2}\,,\qquad F_{\2,2}=dA_{\1,2}\,.
\eea
In this three-equal truncation, the three original \(SL(2,\mathbb{R})\) global symmetries are identified, namely, \(S=T=U\). Hence, the truncation has been referred to as the \(S^3\) truncation, as STU becomes \(S^3\). However, \(S^3\) can easily be confused with a 3-sphere, so we shall not adopt this terminology and instead call it the three-equal truncation.

It is worth pointing out that one can further set \(A_{\1,1}=0\) consistently, provided that the axion is also set to zero. We shall discuss this case below. However, it is not consistent to set \(A_{\1,2}=0\) while keeping \(A_{\1,1}\) nonzero.

\subsubsection{$U(1)^2$ truncation}

A two-charge rotating black hole was constructed in \cite{Chow:2010fw}. The corresponding theory actually admits a consistent truncation at the bosonic level. The consistency of the truncation follows from the fact that the bosonic part of the theory is simply a circle reduction of five-dimensional bosonic string theory. The four-dimensional theory contains two \(U(1)\) gauge fields, two dilatons and one axion. Starting from \eqref{lag}, the truncation can be achieved by taking
\bea
&&A_{\1,1}^{\mathrm{STU}}=\mathcal{A}_{\1}^{1,\mathrm{STU}}=0\,,\qquad \chi_2^{\mathrm{STU}}=\chi_3^{\mathrm{STU}}=0\,,\qquad \mathcal{A}_{\1}^{2,\mathrm{STU}}=A_{\1,1}\,,\qquad A_{\1,2}^{\mathrm{STU}}=A_{\1,2}\,,\cr
&&\varphi_2^{\mathrm{STU}}=\varphi_3^{\mathrm{STU}}=\frac{\varphi_2}{\sqrt{2}}\,,\qquad \varphi_1^{\mathrm{STU}}=\varphi_1\,,\qquad \chi_1^{\mathrm{STU}}=\chi\,.
\eea
After this truncation, the reduced Lagrangian is
\bea
\mathcal{L}&=&\Big(R+g^2\big(2\cosh\varphi_1+4\cosh\frac{\varphi_2}{\sqrt{2}}+\chi^2e^{\varphi_1}
\big)\Big)\,{*\oneone}-\frac{1}{2}{*d}\varphi_1\wedge d\varphi_1-\frac{1}{2} {*d}\varphi_2\wedge d\varphi_2\cr
&&-\frac{1}{2}e^{2\varphi_1}{*d}\chi\wedge d\chi
-\frac{1}{2}e^{-\varphi_1-\sqrt{2}\varphi_2} {*F}_{\2,1}\wedge F_{\2,1} -\frac{1}{2}e^{-\varphi_1+\sqrt{2}\varphi_2} {*F}_{\2,2}\wedge F_{\2,2}\cr
&&-\chi F_{\2,1}\wedge F_{\2,2}\,,\label{Two charges truncation action}
\eea
where $F_{\2,i}=dA_{\1,i}$ for $i=1,2$. The theory can also be obtained by setting two of the \(U(1)\) gauge fields to zero; we therefore refer to this truncation as the \(U(1)^2\) truncation.

\subsubsection{$U(1)$ truncation: AdS Kaluza-Klein theory}

The minimal nontrivial theory obtained through consistent truncation is the AdS KK theory, which retains only one \(U(1)\) gauge field and one dilaton. This can be achieved by a further truncation
\bea
\chi=0\,,\qquad A_{\1,1}=0\,,\qquad A_{\1,2}=A_{\1}\,,
\eea
on the three-equal truncation theory \eqref{S3 action}. Alternatively, the AdS KK theory can also be obtained from the $U(1)^2$ truncation \eqref{Two charges truncation action} via the following additional truncation
\be
A_{\1,1}=A_{\1}^{\mathrm{KK}}\,,\qquad A_{\1,2}=0\,,\qquad \chi=0\,,\qquad
 \varphi_1=-\frac{\varphi^{\mathrm{KK}}}{\sqrt{3}}\,,\qquad \varphi_2=-\sqrt{\frac{2}{3}}\varphi^{\mathrm{KK}}\,.
\ee
The final result is the same, and the Lagrangian of the AdS Kaluza-Klein theory is
\be
\mathcal{L}=\Big(R+6g^2\cosh\frac{\varphi}{\sqrt{3}}\Big)\, {*\oneone} -\frac{1}{2} {*d}\varphi\wedge d\varphi-\frac{1}{2}e^{\sqrt{3}\varphi} {*F}_\2\wedge F_\2\,, \qquad F_{\2}=dA_{\1}\,.\label{gauged KK EMD}
\ee
When the constant $g$ vanishes, the theory is simply the original KK theory in four dimensions.

\bigskip
\noindent{\bf A short summary}
\medskip

The gauged STU supergravity theory is the \(U(1)^4\) Cartan truncation of four-dimensional \(SO(8)\) gauged maximal supergravity. The theory admits four-charge, purely electric static AdS black holes, whose uplifts describe the near-horizon geometries of rotating membranes in eleven dimensions \cite{Cvetic:1999xp}. Labelling the \(U(1)^4\) gauge fields as
\be
A_1=\mathcal{A}_{\1}^2\,,\qquad A_2=A_{\1,2}\,,\qquad A_{\3}=\tilde{A}_{\1,1}\,,\qquad A_4=\tilde{\mathcal{A}}_{\1}^1\,,
\ee
they enter the $S^7$ of the membrane solution as follows,
\be
ds_7^2 = \frac{4}{g^2}\sum_{i=1}^4\Big[ d\mu_i^2 + \mu_i^2 \big(d\phi_i +\frac{g}{2} A_i\big)^2\Big]\,,\qquad \hbox{with}\qquad
\mu_1^2 + \mu_2^2 + \mu_3^2 + \mu_4^2=1\,,
\ee
where for simplicity, we set all the scalars to zero. The various truncations that we have discussed can be summarized as follows
\begin{itemize}

\item Pairwise-equal truncation: $A_1=A_2$, $A_3=A_4$, a complex scalar. Supersymmetric

\item EMDA truncation: $A_1=A_2$, $A_3=A_4=0$, a complex scalar. Supersymmetric

\item Three-equal truncation: $A_1$, $A_2=A_3=A_4$, a complex scalar. Supersymmetric

\item $U(1)^2$ truncation: $A_1, A_2$, $A_3\!=\!A_4\!=\!0$, 2 dilatons and one axion. Non-Supersymmetric

\item $U(1)$ (KK) truncation: $A_1$, $A_2=A_3=A_4=0$, a dilaton. Non-supersymmetric

\end{itemize}
These special consistent truncations allow one to study the STU gauged supergravity model in a simpler subsector.

\section{Eight-charge rotating dyonic AdS black holes}
\label{sec:8charge}

We now present the exact solution for the general eight-charge rotating AdS black holes in gauged STU supergravity. As was discussed, the eight-charge solution can be generated from the $SO(3)^3$ rotation of the five-charge seed. However, the five-charge seed is not necessarily simpler as it singles out one gauge field that carries both electric and magnetic charges, while the remaining three gauge carries either electric or magnetic charges. This seed breaks the symmetry among the different gauge fields; furthermore, it makes it more difficult to perform various truncations discussed in the previous sections. For example, it is not possible to perform the pairwise-equal truncation on the five-charge seed directly. For this reason, we shall present the eight-charge solution directly, without bothering with the $SO(3)^3$ symmetry.

The general solution thus contains 10 nontrivial integration constants and can be viewed as a charged generalization of the Kerr-AdS black hole, with parameters \((\mu,a)\) associated with the mass and angular momentum of the neutral rotating black hole. The eight charges are the four electric and four magnetic charges associated with the \(U(1)\) gauge fields. The solution contains no independent scalar-hair parameters. The general 10-parameter solution is very complicated. To present it in a readable and compact form, we first introduce some shorthand notation, largely following the conventions of \cite{Chong:2004na,Ma:2026osr}:
\bea
&&s_i=\sinh\delta_i\,,\qquad c_i=\cosh\delta_i\,,\qquad \hat{s}_i=\sinh\gamma_i\,,\qquad \hat{c}_i=\cosh\gamma_i\,,\qquad i=1,2,3,4,\\
&&c_{i_1\cdots i_n}=c_{i_1}\cdots c_{i_n}\,,\quad s_{i_1\cdots i_n}=s_{i_1}\cdots s_{i_n} \,,\quad \hat{c}_{i_1\cdots i_n}=\hat{c}_{i_1}\cdots \hat{c}_{i_n}\,,\quad \hat{s}_{i_1\cdots i_n}=\hat{s}_{i_1}\cdots \hat{s}_{i_n}\,.\nn
\eea
In other words, in addition to the parameters \((\mu,a)\) of the Kerr metric, we introduce eight parameters, \((\delta_i,\gamma_i)\), to parameterize the eight charges.

Following \cite{Chong:2004na}, we further introduce barred indices according to the following rules: \(\bar{i}\) denotes the three indices remaining after removing \(i\) from the set \(\{1,2,3,4\}\), and \(\overline{ij}\) denotes the two indices remaining after removing \(i\) and \(j\) from the set. For convenience, we also define the following shorthand constants in terms of the notation introduced above:
\bea
m_1&=&1+\frac{1}{2}\sum_i(s_i^2+\hat{s}_i^2)-\sum_is_i^2\hat{s}_i^2+
\frac{1}{2}\Big[\sum_is_i^2\Big]\Big[\sum_i\hat{s}_i^2\Big]\,,\cr
m_2&=&\sum_is_ic_i(\hat{s}_i\hat{c}_{\bar{i}}-\hat{c}_i\hat{s}_{\bar{i}})\,,\qquad
n_1=\sum_i\hat{s}_i\hat{c}_i(c_is_{\bar{i}}-s_ic_{\bar{i}})\,,\cr
n_2&=&(c_{1234}-s_{1234})(\hat{c}_{1234}-\hat{s}_{1234})+
\sum_{i<j}s_{ij}c_{\overline{ij}}(\hat{c}_{ij}\hat{s}_{\overline{ij}}
-\hat{s}_{ij}\hat{c}_{\overline{ij}})\,,\cr
\mathcal{D}_0&=&c_{1234}\hat{s}_{1234}+s_{1234}\hat{c}_{1234}
+\sum_{i<j}s_{ij}c_{\overline{ij}}\hat{s}_{ij}\hat{c}_{\overline{ij}}\,,\cr
\mathcal{C}_0&=&1+\sum_i\big(s_i^2+\hat{s}_i^2+2s_i^2\hat{s}_i^2\big)
+\sum_{i\neq j}s_i^2\hat{s}_j^2+\Big[\sum_is_i^2\Big]\Big[\sum_{j<k}\hat{s}_{jk}^2\Big]
+\Big[\sum_i\hat{s}_i^2\Big]\Big[\sum_{j<k}s_{jk}^2\Big]\cr
&&+\sum_{i<j}\Big[
s_{ij}^2+\hat{s}_{ij}^2+2s_{ij}^2(\hat{s}_{ij}^2+\hat{s}_{\overline{ij}}^2)
+2s_{ij}c_{ij}(\hat{s}_{ij}\hat{c}_{ij}+\hat{s}_{\overline{ij}}\hat{c}_{\overline{ij}})
\Big]-n_1^2-n_2^2\,,\cr
\mathcal{N}&=&\sqrt{n_1^2+n_2^2}\,,\qquad v=-\frac{2\mu}{\mathcal{N}}(m_1n_1+m_2n_2)\,,\qquad
\Xi=1-a^2 g^2\,.\label{8 charges metric constants}
\eea
Based on the above constants, we can define the functions appearing in the metric
\bea
u &=& a\cos\theta\,,\qquad \Delta_\theta=1-g^2(u^2+2uv)\,,\qquad R=\Big(r-\frac{\mu n_2}{\mathcal N}\Big)^2-\mu^2\,,\cr
K&=&2R+\frac{4\mu}{\mathcal N}(n_2m_1-n_1m_2)r
   -\frac{2\mu n_1}{\mathcal N}v+4\mu^2\mathcal C_0\,,\qquad L=2\mu\big(\mathcal N r+2\mu\mathcal D_0\big)\,,\cr
P&=&R(K-R)+L^2\,,\qquad F=P+a^2K+a^4\,,\qquad W^2=P+u^2K+u^4+2uv(R+u^2)\,,\cr
\Delta_r&=&\Xi R+a^2+g^2(P+a^2K)\,.\label{8 charges metric functions}
\eea
Note that the function $\Delta_\theta$ depends only on $\theta$ or $u$, the functions $(R,K,L,P,F,\Delta_r)$ depend only on $r$, and $W$ depends on both $(r,\theta)$. Another fact worth noting is that the gauge coupling $g$ enters only through the constant $\Xi$ and functions $\Delta_\theta$ and $\Delta_r$. Thus, the modifications to the ungauged solution due to gauging are limited. In terms of these functions, the metric can be written in a compact form
\bea
ds_4^2&=&-\frac{\Delta_\theta}{\Xi W}\Big[
R+u^2+g^2\big(P+u^2(K-R)\big)
\Big]dt^2-\frac{2aL\Delta_\theta\sin^2\theta}{\Xi W}dtd\phi\cr
&&+\frac{\sin^2\theta}{\Xi^2W}\Big[\Delta_\theta F-(a^2-u^2-2uv)\Delta_r\Big]d\phi^2+W\Big(
\frac{dr^2}{\Delta_r}+\frac{d\theta^2}{\Delta_\theta}
\Big)\,.\label{8 charges metric}
\eea
The solution contains 10 integration constants: \((\mu,a,\delta_1,\delta_2,\delta_3,\delta_4,\gamma_1,\gamma_2,\gamma_3,\gamma_4)\). If we set all eight charge parameters to zero, i.e., \(\delta_i=0=\gamma_i\), the solution simply reduces to the Kerr-AdS metric. If we further set the mass parameter \(\mu=0\), the metric reduces to the AdS vacuum in nonrotating spheroidal coordinates
\bea
ds^2=
-\frac{\Delta_\theta}{\Xi}(1+g^2r^2)dt^2
+\frac{r^2+u^2}{(r^2+a^2)(1+g^2r^2)}dr^2
+\frac{r^2+u^2}{\Delta_\theta}d\theta^2
+\frac{r^2+a^2}{\Xi}\sin^2\theta d\phi^2\,.
\eea
The above spheroidal coordinates are related to global AdS coordinates through a coordinate transformation, and hence the metric also describes the maximally symmetric AdS vacuum.

For the four gauge fields, the explicit forms of their solutions depend on the form of the Lagrangian \(\mathcal{L}_{\rm K}\). To treat all formulations on an equal footing and present the solutions corresponding to each form of \(\mathcal{L}_{\rm K}\) discussed in the previous section, we introduce some useful shorthand notation. We first define a set of constants
\bea
\alpha_i&=&(s_ic_{\bar{i}}-c_is_{\bar{i}})(\hat{c}_{1234}-\hat{s}_{1234})+\sum_{j\neq i}(c_is_jc_{\overline{ij}}-s_ic_js_{\overline{ij}})(\hat{c}_{ij}\hat{s}_{\overline{ij}}-\hat{s}_{ij}\hat{c}_{\overline{ij}})\,,\cr
\beta_i&=&\hat{s}_i\hat{c}_i(s_{1234}-c_{1234})+\sum_{j\neq i}\hat{s}_j\hat{c}_j(c_{ij}s_{\overline{ij}}-s_{ij}c_{\overline{ij}})\,,\cr
\mathcal{D}_i&=&s_ic_{\bar{i}}\hat{s}_{1234}+c_is_{\bar{i}}\hat{c}_{1234}+\sum_{j\neq i}\big[
c_is_jc_{\overline{ij}}\hat{s}_{ij}\hat{c}_{\overline{ij}}+s_ic_js_{\overline{ij}}\hat{c}_{ij}\hat{s}_{\overline{ij}}
\big]\,,\qquad \tilde{\mathcal{D}}_i=\hat{s}_i\hat{c}_{\bar{i}}s_i^2-\hat{c}_i\hat{s}_{\bar{i}}c_i^2\,,\cr
T_{ij}&=&2(1+2s_i^2)\Bigg\{\hat{s}_j\hat{c}_j\Big[2\hat{s}_{\bar{j}}^2+\sum_{k\neq j}\hat{s}_{\overline{jk}}^2\Big]-(1+2\hat{s}_j^2)\hat{c}_{\bar{j}}\hat{s}_{\bar{j}}
\Bigg\}+2s_i^2\hat{s}_j\hat{c}_j\Big[1+\sum_k\hat{s}_k^2\Big]\,,\cr
\mathcal{C}_i&=&2s_ic_i\Bigg[1+2\hat{s}_i^2+\sum_{j\neq i}\Big(s_j^2+\hat{s}_j^2+2s_j^2(\hat{s}_{ij}^2+\hat{s}_{\overline{ij}}^2)\Big)+\sum_{j<k}\hat{s}_{jk}^2
+\Big[\sum_k\hat{s}_k^2\Big]\Big[\sum_{j\neq i}s_j^2\Big]
\Bigg]\cr
&&+2(c_i^2+s_i^2)\sum_{j\neq i}s_jc_j(\hat{s}_{ij}\hat{c}_{ij}+\hat{s}_{\overline{ij}}\hat{c}_{\overline{ij}})-2n_1\beta_i-2n_2\alpha_i\,,\cr
\tilde{\mathcal{C}}_i&=&(s_{1234}-c_{1234})T_{ii}+\sum_{j\neq i}(s_{ij}c_{\overline{ij}}-c_{ij}s_{\overline{ij}})T_{ij}+2\hat{s}_i\hat{c}_is_{1234}\Big[2+\sum_k\hat{s}_k^2\Big]\cr
&&+2\sum_{j\neq i}\hat{s}_j\hat{c}_j\Big[
s_{ij}c_{\overline{ij}}(\hat{s}_i^2+\hat{s}_j^2)-c_{ij}s_{\overline{ij}}\sum_{k\notin\{i,j\}}\hat{s}_k^2
\Big]\,,\cr
\lambda_i&=&-\frac{1}{\mathcal N}\Big[
 n_1s_ic_i\big(1-2\hat s_i^2+\sum_j\hat s_j^2\big)+n_2(c_i^2+s_i^2)
  (\hat s_i\hat c_{\bar i}-\hat c_i\hat s_{\bar i})\Big]\,.\label{8 charges gauge field constants}
\eea
In addition, we introduce eight parameters $(q_i,p_i)$:
\be
q_i=\frac{1}{\mathcal N}\Big[
 n_2s_ic_i\big(1-2\hat s_i^2+\sum_j\hat s_j^2\big)-n_1(c_i^2+s_i^2)
  (\hat s_i\hat c_{\bar i}-\hat c_i\hat s_{\bar i})\Big]\,,\qquad p_i = \frac{n_1\alpha_i-n_2\beta_i}{\mathcal N}\,.\label{8pqs}
\ee
As we shall see later, these parameters are significant, as they are proportional to the four electric and four magnetic charges carried by the four \(U(1)\) gauge fields. However, we shall not specify which are electric and which are magnetic charges, as the electromagnetic nature of each charge depends on which form of the Lagrangian \(\mathcal{L}_{\rm K}\) is considered.

By means of these constants, we define four sets of linear functions of coordinates $(r,u)$:
\bea
\mathcal{V}_i&=&q_i r+\lambda_i\big(u-\frac{\mu n_1}{\mathcal N}\big)+\mu\mathcal C_i\,,\qquad
\mathcal{L}_i=\frac{n_2\alpha_i+n_1\beta_i}{\mathcal N}r+2\mu\mathcal D_i-p_i\big(u-\frac{\mu n_1}{\mathcal N}\big)\,,\cr
\tilde{\mathcal{V}}_i&=&p_i r+\frac{n_2\alpha_i+n_1\beta_i}{\mathcal N}\big(u-\frac{\mu n_1}{\mathcal N}\big)+\mu\tilde{\mathcal C}_i\,,\qquad
\tilde{\mathcal{L}}_i=-\lambda_i r-2\mu\tilde{\mathcal D}_i+q_i\big(u-\frac{\mu n_1}{\mathcal N}\big)\,.\label{8 charges linear}
\eea
It is worth pointing out that all these functions are independent of the gauge coupling \(g\); they can be taken directly from the solutions of the ungauged theory. To express the four gauge potentials in a more compact form that is suitable for all three forms of \(\mathcal{L}_{\rm K}\), we define two 1-forms
\be
\omega_1=\Delta_\theta(R+u^2)dt+aL\sin^2\theta d\phi\,,\quad
\omega_2=\Delta_\theta Ldt-a(K-R+u^2+2uv)\sin^2\theta d\phi\,.
\ee
At this stage, the gauge coupling \(g\) begins to enter the picture. We can now introduce two auxiliary electromagnetic potentials in a compact form
\bea
E_i&=&\frac{2\mu}{\sqrt{\Xi}W^2}(\mathcal{V}_i\omega_1+\mathcal{L}_i\omega_2)
+\frac{2\mu}{\sqrt{\Xi}}(g^2u\lambda_idt+p_i\cos\theta d\phi)\,,\cr
M_i&=&\frac{2\mu}{\sqrt{\Xi}W^2}(\tilde{\mathcal{V}}_i\omega_1
+\tilde{\mathcal{L}}_i\omega_2)+\frac{2\mu}{\sqrt{\Xi}}\Big[
g^2u\frac{n_2\alpha_i+n_1\beta_i}{\mathcal N}dt-q_i\cos\theta d\phi
\Big]\,,
\eea
and the four \(U(1)\) fields can then be given in terms of this set of auxiliary 1-form potentials, depending on which ${\cal L}_{\rm K}$ is used:
\bea
T^2:&&\qquad A_{\1,1}=M_1\,,\qquad A_{\1,2}=E_2\,,\qquad \mathcal{A}_{\1}^1=M_3\,,\qquad \mathcal{A}_{\1}^2=E_4\,,\nn\\
S^1:&&\qquad  A_{\1,1}=M_1\,,\qquad \tilde A_{\1,2}=-M_2\,,\qquad \mathcal{A}_{\1}^1=M_3\,,\qquad \mathcal{A}_{\1}^2=E_4\,,\nn\\
U(1)^4:&&\qquad  \tilde A_{\1,1}=E_1\,,\qquad A_{\1,2}=E_2\,,\qquad \tilde{\cal{A}}_{\1}^1=E_3\,,\qquad \mathcal{A}_{\1}^2=E_4\,.
\eea
Recall that the Lagrangians corresponding to the three sets of gauge fields above are given in \eqref{lag}, \eqref{lag2} and \eqref{gauged STU dual}, respectively.

Before writing down there pairs of dilatons and axions, we define a permutation operator \(\mathscr{P}_{ij}\), whose effect is to simultaneously interchange $i$ and $j$ indices, namely
\be
\mathscr{P}_{ij}:\quad (s_i,c_i,\hat{s}_i,\hat{c}_i)\leftrightarrow(s_j,c_j,\hat{s}_j,\hat{c}_j)\,.
\ee
We need likewise to define a series of constants
\bea
\eta_{11}&=&s_1^2+s_3^2+\hat{s}_2^2+\hat{s}_4^2+(s_1^2+s_3^2)(\hat{s}_2^2
+\hat{s}_4^2)+(s_1^2-s_3^2)(\hat{s}_3^2-\hat{s}_1^2)\,,\cr
\eta_{12}&=&2\sum_{j\in\{1,3\}}s_jc_j
(\hat{c}_j\hat{s}_{\bar{j}}-\hat{s}_j\hat{c}_{\bar{j}})\,,\cr
\eta_{13}&=&(s_{13}\hat{c}_{24}+c_{13}\hat{s}_{24})^2
+(s_1c_3\hat{s}_1\hat{c}_3+s_3c_1\hat{s}_3\hat{c}_1)^2\,,\cr
\mathcal{X}&=&(c_{13}\hat{s}_{24}+\hat{c}_{24}s_{13})(c_{24}\hat{c}_{13}+\hat{s}_{13}s_{24})
+(s_2c_4\hat{s}_4\hat{c}_2+s_4c_2\hat{s}_2\hat{c}_4)(s_1c_3\hat{s}_1\hat{c}_3
+s_3c_1\hat{s}_3\hat{c}_1)\,,\cr
\xi_{11}&=&(1-\mathscr{P}_{12}\mathscr{P}_{34})(1+\mathscr{P}_{13})
\big[\hat{s}_1\hat{c}_1(s_{\bar{3}}c_3-c_{\bar{3}}s_3)\big]\,,\qquad
\xi_{12}=\mathscr{P}_{12}\mathscr{P}_{34}\mathcal{X}-\mathcal{X}\,,\cr
\xi_{13}&=&(1-\mathscr{P}_{12}\mathscr{P}_{34})\big[(s_{13}\hat{c}_{24}
+c_{13}\hat{s}_{24})(s_2c_4\hat{s}_2\hat{c}_4+s_4c_2\hat{s}_4\hat{c}_2)
\big]\,.\label{8 charges scalar constants}
\eea
The remaining coefficients can be obtained by applying the above permutations to the results given above. For \(k=1,2,3\), we have
\bea
\eta_{2k}=\mathscr{P}_{12}\eta_{1k}\,,\qquad \xi_{2k}=-\mathscr{P}_{12}\xi_{1k}\,,\qquad\eta_{3k}=\mathscr{P}_{23}\eta_{1k}\,,\qquad \xi_{3k}=\mathscr{P}_{23}\xi_{1k}\,.
\eea
With these constants at hand, we can now define the numerators and denominators of the dilatons and axion. For \(A=1,2,3\), we have
\bea
h_A&=&r^2+\Big(u-\frac{\mu n_1}{\mathcal N}\Big)^2+\frac{2\mu}{\mathcal N}\Big[
 \big(n_2r-n_1u+\frac{\mu n_1^2}{\mathcal N}\big)\eta_{A1}
 +\big(n_1r+n_2u-\frac{\mu n_1n_2}{\mathcal N}\big)\eta_{A2}
 +2\mu\mathcal N\eta_{A3}\Big],\cr
f_A&=&\frac{2\mu}{\mathcal N}\Big[
 \big(n_2r-n_1u+\frac{\mu n_1^2}{\mathcal N}\big)\xi_{A1}
 +\big(n_1r+n_2u-\frac{\mu n_1n_2}{\mathcal N}\big)\xi_{A2}
 +2\mu\mathcal N\xi_{A3}\Big]\,.\label{8 charges scalar functions}
\eea
In this way, the three dilatons and three axions can be written as
\be
e^{\varphi_A}=\frac{h_A}{W}\,,\qquad \chi_A=\frac{f_A}{h_A}\,.
\ee
Remarkably, there is no gauge coupling $g$ dependence in solutions of all the scalar fields.

This completes our presentation of the exact solution describing the general eight-charge rotating AdS black holes. While we have carefully checked these expressions, their complexity makes it difficult for us to guarantee that they are entirely free of typos. We therefore provide an accompanying Mathematica file in the arXiv source files, in which all the equations are verified.

We shall study the global structure and black hole thermodynamics in Section \ref{sec:thermo}. However, one important observation is worth mentioning already at this stage. It is straightforward to verify that at the north and south poles, \(\theta=0,\pi\), the metric is degenerate, and the corresponding Euclidean surface gravity associated with the degenerate Killing vector \(\partial/\partial_\phi\) is
\be
\kappa_{\rm N}=\sqrt{1-\frac{2ag^2v}{\Xi}}\,,\qquad \kappa_{\rm S}=\sqrt{1+\frac{2ag^2v}{\Xi}}\,,\qquad \kappa_{\rm N}^2+\kappa_{\rm S}^2=2\,.
\ee
We can rescale $\phi$ so that the resulting angle has the standard period $\Delta\phi=2\pi$, ensuring smoothness at either north or south poles, but no single rescaling can achieve this at both poles. We therefore define the deficit angle at the north and south poles as \(\delta=2\pi(1-\kappa)\), i.e.
\be
\delta_{\rm N}=2\pi\Big[1-\sqrt{1 -\frac{2ag^2v}{\Xi}}\Big]\,,\qquad \delta_{\rm S}=2\pi\Big[
1-\sqrt{1+\frac{2ag^2v}{\Xi}}\Big]\,.\label{delta NS}
\ee
Thus for the general rotating AdS black holes in the gauged STU supergravity model with $a,g \neq0$, there exists a conical singularity associated with the Killing vector $\partial/\partial_\phi$ when $v\ne 0$. The quantity $v$ is given in the last line of \eqref{8 charges metric constants}. As it is a very important parameter, we give it here again
\be
v=-\frac{2\mu}{\mathcal{N}}(m_1n_1+m_2n_2)\,.\label{constant v}
\ee
Thus, for a rotating solution to be free of the conical singularity, it can carry at most seven independent charges, and the corresponding minimal seed for the most general solution has at most four independent charges.

In the next several sections, we shall consider reduced solutions associated with various consistent truncations of the gauged STU theory discussed in Section \ref{sec:theory}. At the level of solutions, implementing a truncation involves more than simply imposing special relations among the charge parameters \((\delta_i,\gamma_i)\), as required by the truncated theories. This is because, even at the level of the \(U(1)^4\) Lagrangian, the four gauge fields do not appear symmetrically or on an equal footing. The four gauge fields are in the equal footing only when all the axions are turned off and the theory reduces to the generalized EMD theory \eqref{genEMD}. As we shall see later, the parametrization in terms of the truncated parameters \((\delta_i,\gamma_i)\) may no longer be the most convenient one for the reduced solutions. This makes it necessary to present the reduced solutions in a simpler form.

\section{Known special solutions: pairwise-equal truncation}
\label{sec:known}

The general dyonic rotating AdS black holes obtained in the previous section are very complicated, containing 10 nontrivial parameters. One may often be interested only in solutions in a particular limit for a specific purpose. As discussed in Section \ref{sec:theory}, there are many consistent truncations of gauged STU supergravity, including both supersymmetric and non-supersymmetric truncations. The general solution allows us to obtain special solutions in all these truncated theories. A notable class of examples is provided by the pairwise truncations, for which the charged rotating AdS black holes are particularly well understood. The purely electric solutions were constructed in \cite{Chong:2004na}, while the dyonic solutions were constructed in \cite{Chow:2013gba}. For completeness, we also present these special solutions here using our notation. This allows us to see explicitly how the most general solution reduces to these special cases. Other special solutions will be presented in later sections.

\subsection{Dyonic rotating AdS black holes in pairwise-equal theory}

At the level of solutions, we impose pairwise equality on both the electric and magnetic charge parameters. Thus, the eight charge parameters satisfy
\bea
&&(\delta_1^{\mathrm{STU}},\delta_2^{\mathrm{STU}},\delta_3^{\mathrm{STU}},\delta_4^{\mathrm{STU}})=
(\delta_1',\delta_2',\delta_1',\delta_2')\,,\cr
&&(\gamma_1^{\mathrm{STU}},\gamma_2^{\mathrm{STU}},\gamma_3^{\mathrm{STU}},\gamma_4^{\mathrm{STU}})=
(\gamma_1',\gamma_2',\gamma_1',\gamma_2')\,.\label{pairwise charges}
\eea
This solution is the dyonic generalization of the purely electric solutions in \cite{Chong:2004na}, and the corresponding result has already been reported in \cite{Chow:2013gba}. We use the primed notation as an intermediate convention, which will ultimately be replaced with a more convenient parametrization.

To express the electric and magnetic dual charges of the solution in a fully symmetric notation, we redefine the parameters and coordinates. We first introduce the notation \(C_\pm\) by
\bea
s_\pm'&=&s_1'c_2'\pm c_1's_2'\,,\qquad
c_\pm'=c_1'c_2'\pm s_1's_2'\,,\qquad
\hat{s}_\pm'=\hat{s}_1'\hat{c}_2'\pm \hat{c}_1'\hat{s}_2'\,,\qquad
\hat{c}_\pm'=\hat{c}_1'\hat{c}_2'\pm \hat{s}_1'\hat{s}_2'\,,\cr
C_\pm&=&1+
\big[(c_\pm')^2+(s_\pm')^2\big]
\big[(\hat c_\pm')^2+(\hat s_\pm')^2\big]\,,
\eea
so that the relevant constants in \eqref{8 charges metric constants} can be rewritten as
\be
n_2=\frac12 \big[(c_1')^2+(s_1')^2\big]
 \big[(\hat c_2')^2+(\hat s_2')^2\big]+\frac12
 \big[(c_2')^2+(s_2')^2\big]
 \big[(\hat c_1')^2+(\hat s_1')^2\big],\quad
\mathcal N=\frac12\sqrt{C_+C_-}.
\ee
We therefore have the corresponding charges
\be
q_1'\pm q_2' =\sqrt{\frac{C_\mp}{C_\pm}}\,
s_\pm'c_\pm'\big[(\hat c_\pm')^2+(\hat s_\pm')^2\big]\,,\qquad
p_1'\pm p_2' =\sqrt{\frac{C_\mp}{C_\pm}}\,
\hat s_\pm'\hat c_\pm'.
\ee
This charge parametrization is indeed very complicated. We introduce four new (unprimed) charge parameters and the corresponding shorthand notation:
\be
s_i=\sinh\delta_i\,,\qquad c_i=\cosh\delta_i\,,\qquad \hat{s}_i=\sinh\gamma_i\,,\qquad \hat{c}_i=\cosh\gamma_i\,,\qquad i=1,2\,.\label{pairwise charge parameters}
\ee
The relations between the new unprimed parameters and the original primed charge parameters are given by
\be
\hat s_1\hat c_1\pm s_2c_2
=\frac{2s_\pm'c_\pm'
 }{C_\pm}\big[(\hat c_\pm')^2+(\hat s_\pm')^2\big],\qquad
s_1c_1\mp\hat s_2\hat c_2
=\frac{2\hat s_\pm'\hat c_\pm'}{C_\pm}.
\ee
Accordingly, the mass parameter \(\mu\) and the radial coordinate \(r\) also need to be redefined, namely
\be
\mu^{\mathrm{STU}}=\frac{2\mu}{\sqrt{C_+C_-}}\,,\qquad
r^{\mathrm{STU}}=r-\mu+\frac{4\mu n_2}{C_+C_-}\,.
\ee
We continue to use \(u=a\cos\theta\), and then define the functions \(r_i\) and \(u_i\) as follows
\bea
r_1&=&r+\mu\sum_{i=1}^2(\hat{s}_i^2\hat{c}_i^2-s_i^2c_i^2)\,,\qquad r_2=r-\mu\sum_{i=1}^2(\hat{s}_i^2\hat{c}_i^2-s_i^2c_i^2)\,,\cr
u_1&=&u-2\mu\sum_{i=1}^2\hat{s}_i\hat{c}_is_ic_i\,,\qquad u_2=u+2\mu\sum_{i=1}^2\hat{s}_i\hat{c}_is_ic_i\,.
\eea
It is clear that we have $u_1+u_2=2u$ and $r_1+r_2=2r$. We define new functions
\bea
W&=&r_1r_2+u_1u_2,\qquad \sqrt F=W+a^2\sin^2\theta,\qquad
L=2\mu\Big[r-\mu\sum_{i=1}^2
   \bigl(s_i^2c_i^2+\hat s_i^2\hat c_i^2\bigr)\Big],\cr
\Delta&=&\sqrt F-L,\qquad \Delta_r=\Xi\Delta+g^2F,\qquad
N=F-a^2\sin^2\theta\,\Delta \,,\qquad \Delta_\theta=1-g^2u^2\,.\label{pairwise functions}
\eea
Interestingly, the metric can be written in the exact same form as the one presented in \cite{Ma:2026osr}, namely
\be
ds^2=-\frac{W\Delta_r\Delta_\theta}{\Xi N}dt^2+\frac{N\sin^2\theta}{\Xi W}\Big(d\phi-\frac{a\Delta_\theta L}{N}dt\Big)^2
+W\Big(\frac{dr^2}{\Delta_r}+\frac{d\theta^2}{\Delta_\theta}\Big)\,.\label{pairwise metric}
\ee
The two Maxwell fields are simply
\be
A_{\1,i}=\frac{2\sqrt{2}\mu}{\sqrt{\Xi}W}
\Big[
\big(s_i c_ir_1+\hat s_i\hat c_i\,u_1\big)
\big(\Delta_\theta dt-a\sin^2\theta\,d\phi\big)+\hat s_i\hat c_i\,W\cos\theta
\big(ag^2dt-d\phi\big)\Big]\,.
\ee
The dilaton and axion are
\be
e^{\varphi}=\frac{r_1^2+u_1^2}{W}\,,\qquad \chi=\frac{r_2u_1-r_1u_2}{r_1^2+u_1^2}\,.
\ee
It is important to note that, for the choice of charge parameters in \eqref{pairwise charges} corresponding to the pairwise-equal truncation, \(v\) in \eqref{constant v} vanishes for all configurations of the remaining pairwise-equal electric and magnetic charges, and hence there is no conical singularity associated with the azimuthal angle \(\phi\). This dyonic rotating solution was constructed in \cite{Chow:2013gba}, while the purely electric solution with \(\hat{s}_i=0\) was constructed in \cite{Chong:2004na}.

\subsection{Dyonic rotating AdS black holes in EMDA theory}

The EMDA theory is obtained by further truncating the pairwise-equal theory discussed above. The corresponding solution can be obtained by setting the pairwise-equal parameters in \eqref{pairwise charge parameters} to
\bea
(\delta_1^{\mathrm{PW}},\delta_2^{\mathrm{PW}},\gamma_1^{\mathrm{PW}},\gamma_2^{\mathrm{PW}})=
(0,\delta,0,\gamma)
\eea
and use \(\delta\), \(\gamma\) to define new \(s\), \(c\) and \(\hat{s}\), \(\hat{c}\). The two functions \(W\) and \(L\) in \eqref{pairwise functions} simplify to
\bea
W=r^2+u^2-\mu^2(s^2c^2+\hat s^2\hat c^2)^2,\qquad
L=2\mu\big[r-\mu(s^2c^2+\hat s^2\hat c^2)\big]\,.
\eea
Both the metric and the single complex scalar field retain exactly the same forms, while the remaining Maxwell field simplifies to
\be
A_{\1}=\frac{2\sqrt{2}\mu}{\sqrt\Xi W}\Big[
sc\frac{L}{2\mu}\big(\Delta_\theta dt-a\sin^2\theta d\phi\big)+\hat s\hat c\cos\theta
\big(a(\Xi+g^2\sqrt F)dt-\sqrt Fd\phi\big)
\Big]\,.
\ee
Although this solution is merely a special case of the previous pairwise-equal solution, it is still worth singling out, as it provides a dyonic AdS generalization of the well-studied purely electric Kerr-Sen black hole \cite{Sen:1992ua}, unlike the more general pairwise-equal solution.

\section{Solution I: three-equal truncation}
\label{sec:special1}

We now consider dyonic rotating AdS black holes in other truncations discussed in Section \ref{sec:theory}, and the resulting solutions are generally new. Although they are special cases of the general eight-charge solution, they are nevertheless worth presenting because they are nontrivial yet much simpler than the most general solution. We first consider the three-equal truncation. As explained in Section \ref{sec:theory}, four-dimensional STU supergravity can also be obtained from five-dimensional STU supergravity, whose three gauge fields can be permuted among themselves under the \(\mathbb Z_3\) symmetry. Thus, the four gauge fields in four-dimensional STU supergravity have a natural 3+1 splitting, and we can consistently set the three gauge fields equal. The resulting theory can also be viewed as a circle reduction of five-dimensional minimal supergravity. Consequently, one dilaton and one axion also survive. At the level of solutions, this three-equal truncation is equivalent to setting the charge parameters as
\be
(\delta_1^{\mathrm{STU}},\delta_2^{\mathrm{STU}},\delta_3^{\mathrm{STU}},
\delta_4^{\mathrm{STU}})=(\delta_1,\delta_1,\delta_1,\delta_2)\,,\quad
(\gamma_1^{\mathrm{STU}},\gamma_2^{\mathrm{STU}},\gamma_3^{\mathrm{STU}},
\gamma_4^{\mathrm{STU}})=(\gamma_1,\gamma_1,\gamma_1,\gamma_2)\,.\label{S3 charges parameters}
\ee
Note that we do not use primed notation in this truncation, as the original notation is already the most convenient one. At this point, the constants defined in \eqref{8 charges metric constants} that appear in the metric simplify to
\bea
m_1&=&1+\frac{1}{2}\Big[
3s_1^2+s_2^2+3\hat{s}_1^2+\hat{s}_2^2+3s_1^2\hat{s}_1^2+3s_1^2\hat{s}_2^2+3s_2^2\hat{s}_1^2-s_2^2\hat{s}_2^2
\Big]\,,\cr
m_2&=&3s_1c_1\hat{s}_1\hat{c}_1(\hat{c}_1\hat{c}_2-\hat{s}_1\hat{s}_2)+s_2c_2(\hat{s}_2\hat{c}_1^3-\hat{c}_2\hat{s}_1^3)\,,\cr
n_1&=&3s_1c_1\hat{s}_1\hat{c}_1(s_1s_2-c_1c_2)+\hat{s}_2\hat{c}_2(s_1^3c_2-c_1^3s_2)\,,\cr
n_2&=&(c_1^3c_2-s_1^3s_2)(\hat{c}_1^3\hat{c}_2-\hat{s}_1^3\hat{s}_2)
+3s_1c_1\hat{s}_1\hat{c}_1(s_1c_2-c_1s_2)(\hat{c}_1\hat{s}_2-\hat{s}_1\hat{c}_2)\,,\cr
\mathcal{D}_0&=&c_1^3c_2\hat{s}_1^3\hat{s}_2+s_1^3s_2\hat{c}_1^3\hat{c}_2
+3s_1c_1\hat{s}_1\hat{c}_1(s_1\hat{s}_1c_2\hat{c}_2+c_1\hat{c}_1s_2\hat{s}_2)\,,\cr
\mathcal{C}_0&=&\big[1+3s_1^2+s_2^2+3s_1^2(s_1^2+s_2^2)\big]\big[1+3\hat{s}_1^2+\hat{s}_2^2+3\hat{s}_1^2(\hat{s}_1^2+\hat{s}_2^2)\big]
-3s_1^2\hat{s}_1^2(s_1^2+s_2^2)(\hat{s}_1^2+\hat{s}_2^2)\cr
&&+3s_1^2\hat{s}_1^2+s_2^2\hat{s}_2^2+6s_1c_1\hat{s}_1\hat{c}_1(s_1c_1+s_2c_2)(\hat{s}_1\hat{c}_1+\hat{s}_2\hat{c}_2)-n_1^2-n_2^2\,,\cr
\mathcal{N}&=&\sqrt{n_1^2+n_2^2}\,,\qquad v=-\frac{2\mu}{\mathcal{N}}(m_1n_1+m_2n_2)\,.
\eea
The metric functions and the metric itself take the exact same form the general eight-charge case, given by \eqref{8 charges metric functions} and \eqref{8 charges metric}; however the expressions are significantly simpler.

Since only one dilaton and one axion remain, only one set of the scalar constants in \eqref{8 charges scalar constants} is left, given by
\bea
\eta_1&=&2s_1^2+(1+2s_1^2)(\hat{s}_1^2+\hat{s}_2^2)\,,\quad
\eta_2=4s_1c_1\hat{s}_1\hat{c}_1(\hat{s}_1\hat{s}_2-\hat{c}_1\hat{c}_2)\,,\cr
\eta_3&=&(s_1^2\hat{c}_1\hat{c}_2+c_1^2\hat{s}_1\hat{s}_2)^2+4s_1^2c_1^2\hat{s}_1^2\hat{c}_1^2\,,\cr
\xi_1&=&-\hat{s}_1\hat{c}_1(s_1^3c_2-c_1^3s_2)+(2\hat{s}_1\hat{c}_1-\hat{s}_2\hat{c}_2)s_1c_1(s_1s_2-c_1c_2)\,,\cr
\xi_2&=&c_1c_2\big[
(\hat{s}_1^2-s_1^2)\hat{c}_1\hat{c}_2-(1+s_1^2+\hat{s}_1^2)\hat{s}_1\hat{s}_2
\big]+s_1s_2\big[(1+s_1^2+\hat{s}_1^2)\hat{c}_1\hat{c}_2+(s_1^2-\hat{s}_1^2)\hat{s}_1\hat{s}_2
\big]\cr
&&+2s_1c_1\hat{s}_1\hat{c}_1(s_1c_2-c_1s_2)(\hat{s}_1\hat{c}_2-\hat{c}_1\hat{s}_2)\,,\cr
\xi_3&=&(s_1^2\hat{c}_1\hat{c}_2+c_1^2\hat{s}_1\hat{s}_2)(s_1c_2\hat{s}_1\hat{c}_2+s_2c_1\hat{s}_2\hat{c}_1)
-2s_1c_1\hat{s}_1\hat{c}_1(s_1s_2\hat{c}_1^2+c_1c_2\hat{s}_1^2)\,.
\eea
We these constants, we find that the dilaton and axion are given by
\be
e^{\frac{\varphi}{\sqrt{3}}}=\frac{W}{h}\,,\qquad \chi=\frac{\sqrt{3}f}{h}\,,
\ee
where
\bea
h&=&r^2+\Big(u-\frac{\mu n_1}{\mathcal N}\Big)^2+\frac{2\mu}{\mathcal N}\Big[
 \big(n_2r-n_1u+\frac{\mu n_1^2}{\mathcal N}\big)\eta_{1}
 +\big(n_1r+n_2u-\frac{\mu n_1n_2}{\mathcal N}\big)\eta_{2}
 +2\mu\mathcal N\eta_{3}\Big],\cr
f&=&\frac{2\mu}{\mathcal N}\Big[
 \big(n_2r-n_1u+\frac{\mu n_1^2}{\mathcal N}\big)\xi_{1}
 +\big(n_1r+n_2u-\frac{\mu n_1n_2}{\mathcal N}\big)\xi_{2}
 +2\mu\mathcal N\xi_{3}\Big]\,.
\eea
Next, we present the two gauge fields. We first note that the constants relevant to the gauge fields in \eqref{8 charges gauge field constants} are simplified to become
\bea
\alpha_1&=&s_1c_1(c_1c_2-s_1s_2)(\hat{c}_1^3\hat{c}_2-\hat{s}_1^3\hat{s}_2)\cr
&&+\hat{s}_1\hat{c}_1(\hat{c}_1\hat{s}_2-\hat{s}_1\hat{c}_2)\big[
(c_1^2+s_1^2)(s_1c_2-c_1s_2)+s_1c_1(c_1c_2-s_1s_2)
\big]\,,\cr
\alpha_2&=&(c_1^3s_2-s_1^3c_2)(\hat{c}_1^3\hat{c}_2-\hat{s}_1^3\hat{s}_2)+3s_1c_1\hat{s}_1\hat{c}_1(s_1s_2-c_1c_2)(\hat{c}_1\hat{s}_2-\hat{s}_1\hat{c}_2)\,,\cr
\beta_1&=&\hat s_1\hat c_1(s_1^3s_2-c_1^3c_2)
 +(2\hat s_1\hat c_1-\hat s_2\hat c_2)s_1c_1(c_1s_2-s_1c_2)\,,\cr
\beta_2&=&\hat s_2\hat c_2(s_1^3s_2-c_1^3c_2)
 +3s_1c_1\hat s_1\hat c_1(s_1c_2-c_1s_2)\,,\qquad p_i=\frac{n_1\alpha_i-n_2\beta_i}{\mathcal N}\,,\cr
q_1&=&\frac{1}{\mathcal N}\Big[
 n_2s_1c_1(1+\hat s_1^2+\hat s_2^2)
 -n_1(c_1^2+s_1^2)\hat s_1\hat c_1
       (\hat c_1\hat c_2-\hat s_1\hat s_2)\Big]\,,\cr
q_2&=&\frac{1}{\mathcal N}\Big[
 n_2s_2c_2(1+3\hat s_1^2-\hat s_2^2)
 -n_1(c_2^2+s_2^2)(\hat s_2\hat c_1^3-\hat c_2\hat s_1^3)\Big]\,,\cr
\lambda_1&=&-\frac{1}{\mathcal N}\Big[
 n_1s_1c_1(1+\hat s_1^2+\hat s_2^2)
 +n_2(c_1^2+s_1^2)\hat s_1\hat c_1
     (\hat c_1\hat c_2-\hat s_1\hat s_2)\Big],\cr
\lambda_2&=&-\frac{1}{\mathcal N}\Big[
 n_1s_2c_2(1+3\hat s_1^2-\hat s_2^2)
 +n_2(c_2^2+s_2^2)(\hat s_2\hat c_1^3-\hat c_2\hat s_1^3)\Big]\,,\cr
\mathcal{D}_1&=&\hat{s}_1\hat{c}_1(s_1^2\hat{c}_1\hat{c}_2-c_1^2\hat{s}_1\hat{s}_2)\,,\qquad
\mathcal{D}_2=s_2c_1^3\hat{s}_1^3\hat{s}_2+c_2s_1^3\hat{c}_1^3\hat{c}_2
+3s_1c_1\hat{s}_1\hat{c}_1(c_2c_1\hat{c}_1\hat{s}_2+s_2s_1\hat{s}_1\hat{c}_2)\,,\cr
\mathcal C_1&=&2\mathcal N\lambda_1+\beta_1+\frac{\big[n_2(2m_1+1+\eta_1)-n_1(2m_2-\eta_2)\big]p_1
 +(n_2\xi_1+n_1\xi_2)(2q_1-q_2)}{\mathcal N}\,,\cr
\mathcal C_2&=&-2(n_2\alpha_2+n_1\beta_2)+\frac{\big[2n_2+3(n_2\eta_1+n_1\eta_2)\big]q_2
 -3(n_2\xi_1+n_1\xi_2)p_1+n_1\lambda_2}{\mathcal N}\,.
\eea
while the linear functions in \eqref{8 charges linear} reduce to
\bea
\mathcal V_1&=&p_1r+
 \frac{n_2\alpha_1+n_1\beta_1}{\mathcal N}
 \big(u-\frac{\mu n_1}{\mathcal N}\big)
 +\mu\mathcal C_1\,,\qquad
\mathcal L_1=-\lambda_1r-2\mu\mathcal D_1
 +q_1\big(u-\frac{\mu n_1}{\mathcal N}\big),\cr
\mathcal V_2&=&q_2r+\lambda_2\big(u-\frac{\mu n_1}{\mathcal N}\big)
 +\mu\mathcal C_2,\qquad
\mathcal L_2=\frac{n_2\alpha_2+n_1\beta_2}{\mathcal N}r
 +2\mu\mathcal D_2-p_2\big(u-\frac{\mu n_1}{\mathcal N}\big)\,.\label{S3 gauged linear}
\eea
After defining two auxiliary 1-forms,
\bea
\omega_1=\Delta_\theta(R+u^2)dt+aL\sin^2\theta d\phi\,,\qquad \omega_2=\Delta_\theta Ldt-a(K-R+u^2+2uv)\sin^2\theta d\phi\,,
\eea
the two Maxwell fields are
\bea
A_{\1,1}&=&\frac{2\sqrt{3}\mu}{\sqrt{\Xi}W^2}(\mathcal{V}_1\omega_1+\mathcal{L}_1\omega_2)
+\frac{2\sqrt{3}\mu}{\sqrt{\Xi}}\Big[g^2u\frac{n_2\alpha_1+n_1\beta_1}{\mathcal{N}}dt-q_1\cos\theta d\phi
\Big]\,,\cr
A_{\1,2}&=&\frac{2\mu}{\sqrt{\Xi}W^2}(\mathcal{V}_2\omega_1+\mathcal{L}_2\omega_2)+\frac{2\mu}{\sqrt{\Xi}}\big[
g^2u\lambda_2dt+p_2\cos\theta d\phi\big]\,.\label{S3 gauge field}
\eea

It is clear that even after the truncation, the four-charge dyonic solution is still very complicated; it becomes much simpler when we set \(\gamma_1=\gamma_2=0\) so that they are purely electric. This is a special three-equal truncation of the four-charge purely-electric solutions of \cite{Ma:2026osr,Wu:2026jwa}. As this specialization was not presented in the literature, we give it here. Following the notation in \cite{Ma:2026osr}, we define $r_i=r+2\mu s_i^2$ and $u=a\cos\theta$\,. Other constants and functions are
\bea
\Xi&=&1-a^2g^2\,,\qquad \alpha_1=s_1c_1(c_{12}-s_{12})\,,\qquad \alpha_2=s_2c_1^3-c_2s_1^3\,,\qquad P=r_1^3r_2\,,\cr
\Delta_\theta&=&1-g^2u^2\,,\qquad \Delta=r^2-2\mu r+a^2\,,\qquad L=2\mu\big[rc_1^3c_2-(r-2\mu)s_1^3s_2\big]\,,\cr
R&=&r^2-2\mu r\,,\qquad
W^2=(r_1^2+u^2)(r_1r_2+u^2)-4\mu^2u^2s_1^2c_1^2(c_1s_2-c_2s_1)^2\,,\cr
F&=&(r_1^2+a^2)(r_1r_2+a^2)-4\mu^2a^2s_1^2c_1^2(c_1s_2-c_2s_1)^2\,,\qquad
\Delta_r=\Xi\Delta+g^2F\,,\cr
N&=&F-a^2\sin^2\theta\Delta\,,\qquad K=r_1(r_1+r_2)-4\mu^2s_1^2c_1^2(c_1s_2-c_2s_1)^2\,.
\eea
The metric has exactly the same form as in the pairwise-equal case \eqref{pairwise metric}. To define the Maxwell fields, we introduce two sets of auxiliary 1-forms
\bea
E_i\!&=&\!\frac{2\mu}{\sqrt{\Xi}r_iW^2}\Bigg\{
\Delta_\theta\Big[s_ic_i\big(P+u^2(K-R)\big)-\alpha_iLu^2\Big]dt-a\sin^2\theta\big[\alpha_i(P+u^2R)+s_ic_iL u^2\big]d\phi
\Bigg\}\,,\cr
M_i\!&=&\!\frac{2\mu\cos\theta}{\sqrt{\Xi}W^2}\Bigg\{
a\Big[\Delta_\theta\big(\alpha_i(R+u^2)+s_ic_iL\big)+g^2\alpha_iW^2\Big]dt-\big[s_ic_iN-a^2\sin^2\theta\alpha_i L\big]d\phi
\Bigg\}\,.
\eea
Then \eqref{S3 gauge field} become
\be
A_{\1,1}=\sqrt{3}M_1\,,\qquad A_{\1,2}=E_2\,.
\ee
The the scalar fields are now simplified to
\be
e^{\frac{\varphi}{\sqrt{3}}}=\frac{W}{r_1^2+u^2}\,,\qquad \chi=\frac{\sqrt{3}f}{r_1^2+u^2}\,,\qquad
f=2\mu u\,s_1c_1(c_1s_2-c_2s_1)\,.
\ee

\section{Solution II: $U(1)^2$ truncation}
\label{sec:special2}

The two-charge (purely electric) solutions were constructed in \cite{Chow:2010fw}. In this section, we present the four-charge dyonic solution in the $U(1)^2$ truncation of the \(U(1)^4\) theory. The reduced theory contains two Maxwell fields, two dilatons and one axion. At the level of solutions, we may take
\bea
&&(\delta_1^{\mathrm{STU}},\delta_2^{\mathrm{STU}},\delta_3^{\mathrm{STU}},\delta_4^{\mathrm{STU}})
=(0,\delta_2',0,\delta_1')\,,\quad
(\gamma_1^{\mathrm{STU}},\gamma_2^{\mathrm{STU}},\gamma_3^{\mathrm{STU}},\gamma_4^{\mathrm{STU}})
=(0,\gamma_2',0,\gamma_1')\,.
\eea
As in the previous examples, we use primed parameters for the same reason: they do not provide the most convenient parametrization of the solution space. To see this explicitly, we present the remaining electric and magnetic charges:
\bea
p_i'&=&
\frac{\hat s_i'\hat c_{\bar i}'}{\mathcal N}
\Big[(c_{\bar i}')^2(\hat c_i')^2
     +(s_{\bar i}')^2(\hat s_{\bar i}')^2\Big]\,,\qquad
q_i'=
\frac{s_i'c_{\bar i}'\hat c_i'\hat c_{\bar i}'
      +c_i's_{\bar i}'\hat s_i'\hat s_{\bar i}'}{\mathcal N}
\Big[(c_i')^2(\hat c_{\bar i}')^2
     +(s_i')^2(\hat s_i')^2\Big]\,,\nn\\
\mathcal N
&=&\sqrt{\Big[(c_1')^2(\hat c_2')^2+(s_1')^2(\hat s_1')^2\Big]
       \Big[(c_2')^2(\hat c_1')^2+(s_2')^2(\hat s_2')^2\Big]}\,,\qquad i=1,2\,.
\eea
We shall introduce a new set of (unprimed) parameters $(\delta_1,\delta_2,\gamma_1,\gamma_2)$, so that the charges $(q_i,p_i)$ are expressed in a symmetric form. Specifically, we define the charges and other related constants as follows
\bea
&&s_i=\sinh\delta_i\,,\qquad c_i=\cosh\delta_i\,,\qquad \hat{s}_i=\sinh\gamma_i\,,\qquad \hat{c}_i=\cosh\gamma_i\,,\cr
&&q_i=s_ic_i\,,\qquad p_i=\hat{s}_i\hat{c}_i\,,\qquad d_i=\frac{1+\sqrt{1+4(q_i^2+p_{\bar{i}}^2)}}{2}\,,\qquad D=\sqrt{d_1d_2}\,,\cr
&&\mathcal{J}_i=q_ip_i-q_{\bar{i}}p_{\bar{i}}\,,\qquad \eta=\sqrt{1-\frac{\mathcal{J}^2_1}{D^2}}\,,\qquad \mathcal{I}=q_1q_2+p_1p_2\,,\qquad
\mathcal{S}=q_1p_1+q_2p_2\,.\label{Two charges truncation constants}
\eea
To establish the relation between the old (primed) and new (unprimed) parameters, we further define the constants associated with the new parameters \(\delta_i,\gamma_i\):
\bea
x_i=\frac{q_i}{\eta}\sqrt{\frac{d_{\bar{i}}}{d_i}}\,,\qquad y_i=\frac{p_i}{\eta}\sqrt{\frac{d_i}{d_{\bar{i}}}}\,.
\eea
The relations between the old and new charge parameters are then given by
\bea
\sinh(\gamma_1'\pm\gamma_2')=y_1\pm y_2\,,\qquad \sinh(\delta_1'\pm\delta_2')=\frac{x_1\pm x_2}{\sqrt{1+(y_1\pm y_2)^2}}\,.
\eea
The transformation of the mass parameter \(\mu\) and the radial coordinate \(r\) is
\bea
\mu^{\mathrm{STU}}=\eta \mu\,,\qquad r^{\mathrm{STU}}=r-\mu+\eta \mu\frac{ n_2}{\mathcal{N}}\,,\qquad
\mathcal{N}=\frac{D}{\eta^2}\,,\qquad n_2=\frac{D}{\eta}-2\mathcal{D}_0\,.
\eea
We continue to impose \(u=a\cos\theta\) and \(\Xi=1-a^2g^2\). In terms of the new coordinates and parameters, we then define new functions and constants
\bea
&&r_i=r+2\mu(d_i-1)\,,\qquad \Delta=(r-\mu)^2-\mu^2\eta^2+a^2\,,\qquad v=\frac{\mu(d_2-d_1)\mathcal{J}_1}{\eta D}\,,\cr
&&H_i(r,u)=rr_{\bar{i}}+u^2+\mu^2(1-\eta^2)+\frac{2\mu d_{\bar{i}}\mathcal{J}_i}{\eta D}u\,,\quad W=\sqrt{H_1H_2}\,,\quad
L=\frac{2\mu D}{\eta}\big[r-\mu(1-\eta^2)\big]\,,\cr
&&P=H_1(r,0)H_2(r,0)\,,\qquad K=H_1(r,0)+H_2(r,0)-\frac{4\mu^2\mathcal{J}_1^2}{\eta^2}\,,\qquad F=P+a^2K+a^4\,,\cr
&&\Delta_\theta=1-g^2(u^2+2uv)\,,\qquad \Delta_r=\Xi\Delta+g^2F\,,\qquad N=F-(a^2\sin^2\theta-2uv)\Delta\,.\label{Two charges truncation functions}
\eea
Here we should note that \(H_i(r,0)\) denotes the value of \(H_i\) at the equator, \(u=0\). The metric then takes the same form as the pairwise case \eqref{pairwise metric}. To keep the gauge potentials in a compact forms, we continue to employ the notation \(\alpha_i\) and \(\lambda_i\) introduced in \eqref{8 charges gauge field constants}, which now take the simpler forms
\bea
\alpha_i=\frac{d_{\bar{i}}q_i+p_i\mathcal{J}_i}{\eta D}\,,\qquad \lambda_i=-\frac{d_{i}p_i+q_i\mathcal{J}_i}{\eta D}\,.
\eea
The two \(U(1)\) fields can be combined and written as
\bea
A_{\1,i}&=&\frac{2\mu}{\sqrt{\Xi}H_{\bar{i}}}\Bigg\{
\Big[\Delta_\theta\big(q_ir+\frac{\mu p_i\mathcal{J}_i}{d_{\bar{i}}}\big)+\lambda_iu(\Delta_\theta+g^2H_{\bar{i}})\Big]dt\cr
&&-\Big[a\sin^2\theta\big[\alpha_ir-p_iu+\frac{\mu \mathcal{J}_i}{\eta D}(d_ip_i+\frac{\mathcal{I}p_{\bar{i}}}{d_i})\big]-p_iH_{\bar{i}}\cos\theta
\Big]d\phi
\Bigg\}\,.
\eea
The scalar functions \eqref{8 charges scalar functions} of the eight-charge case also simplify significantly and only one pair survives; in the new coordinates and parameters, we denote them with the subscript 0:
\bea
h_0&=&r^2+u^2+2\mu\Bigg[\Big(\frac{p_1^2}{d_2}+\frac{p_2^2}{d_1}\Big)r+\Big(\frac{\alpha_1p_1}{d_2}+\frac{\alpha_2p_2}{d_1}\Big)u\Bigg]
+\frac{\mu^2}{D^2}(\mathcal{S}^2+4p_1^2p_2^2)\,,\cr
f_0&=&2\mu\Bigg[
\Big(\frac{q_1p_2}{d_1}+\frac{q_2p_1}{d_2}\Big)r+\Big(\frac{\alpha_1q_2}{d_2}+\frac{\lambda_1p_2}{d_1}\Big)u+\frac{\mu\mathcal{I}\mathcal{S}}{D^2}
\Bigg]\,.
\eea
The two dilatons and one axion can be written as
\be
e^{\varphi_1}=\frac{h_0}{W}\,,\qquad e^{\frac{\varphi_2}{\sqrt{2}}}=\sqrt{\frac{H_1}{H_2}}\,,\qquad \chi=\frac{f_0}{h_0}\,.
\ee

\section{Solution III: AdS Kaluza-Klein truncation}
\label{sec:special3}

The asymptotically flat dyonic KK black hole was first constructed in \cite{Gibbons:1985ac} and is known as the Gibbons-Wiltshire black hole. Its rotating generalization was constructed in \cite{Rasheed:1995zv,Larsen:1999pp} and is known as the Rasheed-Larsen black hole. The AdS KK theory is a consistent truncation of gauged STU supergravity, in which a specific scalar potential is introduced and the vacuum spacetime is AdS with radius \(\ell=1/g\), as can be seen in \eqref{gauged KK EMD}. An electrically charged rotating black hole was constructed in \cite{Chow:2010sf}, while a dyonic static black hole was constructed in \cite{Lu:2013ura}. However, there has hitherto been no dyonic rotating AdS black hole in this simple theory. This solution is particularly worth presenting separately, as the corresponding AdS KK theory \eqref{gauged KK EMD} is relatively simple. Furthermore, the solution captures the essential properties of dyonic AdS black holes in gauged supergravities.

In terms of the solution, we choose the charge parameters of the black hole solution in \eqref{S3 charges parameters} such that $\delta_1=\gamma_1=0$, and $\delta_2=\delta$ and $\gamma_2=\gamma$. We define
\be
\hat{s}=\sinh\delta\,,\qquad \hat{c}=\cosh\delta\,,\qquad s=\sinh\gamma\,,\qquad c=\cosh\gamma\,.\label{S3 to KK parameters}
\ee
The constants reduce to
\bea
m_1 &=& 1+\frac12(s^2+\hat s^2-s^2\hat s^2)\,,
\qquad m_2 = s\hat s\hat c\,,\qquad
n_1=-sc\hat{s}\,,\qquad n_2=c\hat{c}\,,\cr
\mathcal D_0 &=& 0\,,\qquad
 \mathcal C_0 = -s^4\hat s^2\,,\qquad
\mathcal{N}=c\sqrt{1+c^2\hat{s}^2}\,,\qquad v = \frac{\mu s\hat s(s^2-c^2\hat s^2)}{\sqrt{1+c^2\hat s^2}}\,.
\eea
In \cite{Lu:2013ura}, the electric and magnetic charges are symmetrically parameterized by a pair of constants $(\beta_1,\beta_2)$, which originate from the parameters of the $SL(3,\mathbb R)$ Toda system \cite{Lu:2013uia}. It is also useful to define $(\Gamma_1,\Gamma_2)$ in terms of $(\beta_1,\beta_2)$:
\be
\Gamma_i=1-2\beta_i+\beta_i\beta_{\bar{i}}\,,\qquad i=1,2.
\ee
Here, an overbar denotes the other index among \(1,2\), i.e. \(\bar{1}=2,\ \bar{2}=1\). In order to compare with the static solution \cite{Lu:2013ura}, we use the $(\beta_1,\beta_2)$ charge parameters.
We perform a coordinate transformation on the radial coordinate in \eqref{8 charges metric} and give the correspondence between the old and new charge parameters
\bea
\mu\rightarrow \mu\,,\qquad s^2 =\frac{2\beta_1}{\Gamma_1}\,,\qquad \hat{s}^2=\frac{2\beta_2\Gamma_1}{\Gamma_2(1+\beta_1\beta_2)}\,,\qquad
r\rightarrow r-\frac{2\mu\beta_1\beta_2}{1+\beta_1\beta_2}\,,\label{KK EMD sc beta}
\eea
thereby rewriting the solution. Based on these two new parameters, we define the following constants and functions
\bea
v&=&\frac{4\mu(\beta_1-\beta_2)\sqrt{\beta_1\beta_2}}{\Gamma_1\Gamma_2}\,,\qquad \Xi=1-a^2g^2\,,\qquad u=a\cos\theta\,,\qquad
\Delta=r^2-2\mu r+a^2\,,\cr
\Delta_\theta&=&1-g^2(u^2+2uv)\,,\qquad L=\frac{2\mu}{\sqrt{\Gamma_1\Gamma_2}}\big[(1+\beta_1\beta_2)r-2\mu\beta_1\beta_2\big]\,,\cr
F&=&\Big[r^2+a^2+\frac{4\mu\beta_1}{\Gamma_1}\big((1-\beta_2)r+\mu\beta_2\big)\Big]
\Big[r^2+a^2+\frac{4\mu\beta_2}{\Gamma_2}\big((1-\beta_1)r+\mu\beta_1\big)\Big]-\frac{16\mu^2a^2\beta_1\beta_2}{\Gamma_1\Gamma_2}\,,\cr
\Delta_r&=&\Xi\Delta+g^2F\,,\qquad
N=F-(a^2\sin^2\theta-2uv)\Delta\,,\cr
H_1&=&r^2+u^2+\frac{4\mu}{\Gamma_1}\Big[\beta_1(1-\beta_2)r+\mu\beta_1\beta_2
+\sqrt{\beta_1\beta_2}u\Big]\,,\cr
H_2&=&r^2+u^2+\frac{4\mu}{\Gamma_2}\Big[\beta_2(1-\beta_1)r+\mu\beta_1\beta_2
-\sqrt{\beta_1\beta_2}u\Big]\,.\label{KKstuffs}
\eea
The metric, the Maxwell field, and the dilaton can be written as
\bea
ds^2&=&-\frac{\sqrt{H_1H_2}\Delta_r\Delta_\theta}{\Xi N}dt^2+\frac{N\sin^2\theta}{\Xi\sqrt{H_1H_2}}\Big(d\phi-\frac{a\Delta_\theta L}{N}dt\Big)^2
+\sqrt{H_1H_2}\Big(\frac{dr^2}{\Delta_r}+\frac{d\theta^2}{\Delta_\theta}\Big)\,,\cr
A_{\1}&=&\sqrt{\frac{2}{\Xi}}\frac{\mu}{H_2}\Bigg\{
\Big[\frac{\sqrt{\beta_2\Gamma_1}}{\Gamma_2}\Delta_\theta\partial_rH_1
-\frac{2(1-\beta_1\beta_2)}{\Gamma_2}\sqrt{\frac{\beta_1}{\Gamma_1}}u(\Delta_\theta+g^2H_2)
\Big]dt\cr
&&-\Big[
\frac{1-\beta_1\beta_2}{\Gamma_1}\sqrt{\frac{\beta_2}{\Gamma_2}}a\sin^2\theta\partial_rH_1
-\frac{2\sqrt{\beta_1\Gamma_2}}{\Gamma_1}\big(
a(u+v)\sin^2\theta+H_2\cos\theta
\big)
\Big]d\phi
\Bigg\}\,,\cr
\varphi&=&\frac{\sqrt{3}}{2}\log\frac{H_2}{H_1}\,.
\eea
For \(a=0\), this solution reduces to the static case considered in \cite{Lu:2013ura}. When $\beta_2=0$, the solution reduces to the purely-electric rotating solution, constructed in \cite{Chow:2010sf}.

This rotating dyonic KK AdS black hole illustrates the origin of the deficit angle associated with general rotating AdS black holes. The non-vanishing expression of $v$ in the first line of \eqref{KKstuffs} indicates that a single field strength carrying both electric and magnetic charges is responsible for generating the deficit angle. If we wish the rotating black hole in the gauged theory to be free of deficit angles $v=0$, we can consider the purely electric case \(\beta_1=0\), or the purely magnetic case \(\beta_2=0\), or the case where the electric and magnetic charges satisfy the relation \(\beta_1=\beta_2\). Note that when \(\beta_1=\beta_2\), the dilaton does not decouple unless the rotation is turned off. The rotating dyonic KK AdS black hole confirms the earlier statement that the minimal seed for the most general solution without a conical singularity should have at most four charges. It also illustrates that a solution with fewer than four charges is not necessarily free of the conical singularity.

\section{Thermodynamic analysis}
\label{sec:thermo}

\subsection{An overview and subtleties of scalar charge}

In the preceding sections, we presented the exact solution describing the eight-charge non-extremal rotating AdS black holes in the gauged STU theory, together with the rotating solutions in various truncated theories obtained through the corresponding consistent truncations. Here, we study their global structure, focusing on black hole thermodynamics. We also treat the cosmological constant associated with the gauge coupling \(g\) as a thermodynamic variable, so that the black hole thermodynamics includes the conjugate pair of thermodynamic volume \(V\) and pressure \(\mathcal{P}\) \cite{Kastor:2009wy,Cvetic:2010jb}. Furthermore, since the constant \(v\) defined in \eqref{constant v} is generally nonzero, the solutions generally possess a deficit angle associated with the azimuthal coordinate \(\phi\). However, we shall not incorporate the deficit angle into the first law of thermodynamics, and we therefore take the range of integration of the azimuthal angle \(\phi\) to be the usual range \([0,2\pi]\), at the cost of having conical singularities at both the north and south poles when the constant \(v\) in \eqref{constant v} does not vanish. Nevertheless, we observe that, once the period of \(\phi\) is fixed to a specified value, say \(2\pi\), the tensions at the north and south poles, \(\mathcal{T}_{\rm N,\rm S}=\delta_{\rm N,\rm S}/8\pi\), still vary with \(v\) \eqref{delta NS}. The following 1-form variation nevertheless vanishes identically, i.e.,
\be
\kappa_{\rm N}\delta \mathcal{T}_{\rm N}+\kappa_{\rm S}\delta \mathcal{T}_{\rm S}=-\frac{1}{8}\delta(\kappa_{\rm N}^2+\kappa_{\rm S}^2)=0\,.
\ee
Therefore, they do not contribute to the first law.

The most general solution is specified by 10 integration constants, which parameterize the mass \(M\), angular momentum \(J\), four electric charges \(Q_{eI}\), and four magnetic charges \(Q_{mI}\), with \(I=1,2,3,4\). We would naively expect the first law to read
\be
\delta M=T\delta S+\Omega\delta J+\sum_{I=1}^4(\Phi_{eI}\delta Q_{eI}+\Phi_{mI}\delta Q_{mI})+V\delta\mathcal{P}\,.
\ee
However, this turns out not to be true. Gauged STU supergravity is, from the point of view of Einstein gravity with minimally coupled matter, a garden-variety theory, where all the thermodynamic quantities in the above first law can be precisely and unambiguously calculated. The calculation of the mass of such a complicated rotating AdS solution can be somewhat tricky and the best strategy is perhaps the one given in \cite{Chen:2005zj}, based on the Ashtekar, Magnon and Das (AMD) procedure \cite{Ashtekar:1984zz,Ashtekar:1999jx}. It turns out the right-hand side of the above first law is not integrable, even when we treat the cosmological constant as a thermodynamic constant, with $\delta {\cal P}=0$.

This phenomenon was first observed in \cite{Lu:2013ura} for the static dyonic KK AdS black hole, and it is resolved by including scalar contributions to the first law, even though they are not associated with independent charges. To illustrate this, we consider the KK AdS theory as an example, since it is the simplest truncation containing only one dilaton. In global coordinates, a conformally massless scalar such as the dilaton behaves like
\be
\phi \sim \frac{\phi_1}{\rho} + \frac{\phi_2}{\rho^2} + {\cal O}(1/\rho^3)\,.
\ee
This scalar can contribute to the first law by an additional term
\be
Z=\frac{1}{12} g^2 ( \phi_1 \delta\phi_2-2\phi_2 \delta\phi_1)\,.
\ee
This quantity is closely related to the scalar boundary condition on the asymptotic AdS spacetime. It was shown that there are three types of boundary conditions that preserve the full asymptotic AdS conformal symmetry, namely (1) $\phi_1=0$ \cite{Breitenlohner:1982bm}, (2) $\phi_2=0$ \cite{Klebanov:1999tb} and (3) $\phi_2=c \phi_1^2$, where $c$ is a purely numerical constant \cite{Hertog:2004dr}. It can be easily seen that when one of these three conditions is satisfied, the 1-form $Z$ vanishes and the first law retains its conventional structure.  For the static dyonic KK AdS black hole \cite{Lu:2013ura}, one of the three conditions is indeed satisfied when $\beta_1 \beta_2=0$ or $\beta_1=\beta_2$. However, for generic charge parameters $(\beta_1,\beta_2)$, none of the three boundary conditions is satisfied, so the scalar contribution \(Z\) does not vanish and can be expressed as
\be
Z=X \delta  Y\,,
\ee
where $(X,Y)$ were explicitly given in \cite{Lu:2013ura}. The need to introduce an analogous $Z$ term into the first law is quite common for hairy black holes in asymptotic AdS spacetimes, such as those with massive scalar hair \cite{Liu:2013gja,Lu:2014maa} or Proca vector hair \cite{Liu:2014tra}.  The rule is simple: for a field $\Phi$ satisfying two-derivative equations of motion, its asymptotic AdS boundary potentially contains two parameters, such as $(\phi_1,\phi_2)$ discussed above. If both parameters are nonzero and the dimensionless ratio $\phi_2/\phi_1^\alpha$ for suitable $\alpha$ is not a purely numerical constant, then its contribution to the differential 1-form $Z=XdY$ is nontrivial.

The scalar contribution \(Z\) to the first law was also confirmed for the dyonic KK AdS black hole in \cite{Lu:2013ura} using the Iyer-Wald formalism \cite{Wald:1993nt,Iyer:1994ys}. The formalism was also applied to analyse the more general gauged STU supergravity theory, in which an eight-charge static dyonic AdS black hole was constructed \cite{Chow:2013gba}. The upshot is that the first law should be augmented by contributions from the complex scalars and should take the form
\be
\delta M=T\delta S+\Omega\delta J+\sum_{I=1}^4(\Phi_{eI}\delta Q_{eI}+\Phi_{mI}\delta Q_{mI})+V\delta\mathcal{P} + Z\,,\label{firstlaw}
\ee
where
\be
Z=\sum_{A=1}^3(X_{\varphi A}\delta Y_{\varphi A}+X_{\chi A}\delta Y_{\chi A})\,.\label{Zform}
\ee
Note that it was proposed in \cite{Hertog:2004ns} that, within the framework of designer gravity, the quantity \(Z\) can be integrated to give an additional contribution to the mass \(M\), thereby recovering the standard form of the black hole first law. In this paper, we shall not carry out this procedure, as it is straightforward once $(X,Y)$ are known.

\subsection{Eight charge solution}
\label{thermodynamics 8 charges}

First, we consider the thermodynamics of the most general eight-charge solution in Section \ref{sec:8charge}. As mentioned above, the solution is asymptotically AdS in spheroidal coordinates. The horizon is determined by the radial function \(\Delta_r\), which is a quartic polynomial in the radial coordinate \(r\). For sufficiently large \(\mu\), the equation \(\Delta_r=0\) has positive roots. The black hole horizon is the largest root, denoted by \(r_+\), i.e., \(\Delta_r(r_+)=0\). The thermodynamic quantities, such as the mass, entropy, temperature, angular momentum, angular velocity, pressure, and volume, can be expressed in relatively simple forms, given by
\bea
M&=&\frac{\mu}{\Xi\mathcal{N}}(n_2m_1-n_1m_2)\,,\quad S=\frac{\pi}{\Xi}\sqrt{F(r_+)}\,,\quad T=\frac{\Delta_r'(r_+)}{4\pi\sqrt{F(r_+)}}\,,\quad
J=\frac{\mu a}{\Xi}\mathcal{N}\,,\cr
\Omega&=&\frac{a\Xi L(r_+)}{F(r_+)}\,,\qquad \mathcal{P}=\frac{3g^2}{8\pi}\,,\qquad V=\frac{\pi F'(r_+)}{3\Xi}+\frac{4\pi a^2}{3}M\,.\label{8 charges M J}
\eea
Here we should note that, since the spacetime is static at infinity, the angular velocity is evaluated on the horizon.

The calculations of the electric and magnetic charges are both standard. The equations of motion yield four closed 2-forms \(G_I\), which determine the electric charges, while the magnetic charges are obtained from the Bianchi identities. From the Lagrangian \eqref{lag}, we have
\bea
G_1&=&e^{-\varphi_1+\varphi_2-\varphi_3}{*F}_{\2,1}
-\chi_3e^{-\varphi_1+\varphi_2+\varphi_3}{*F}_{\2,2}
+\chi_1\big(\mathcal F_{\2}^1-\chi_3\mathcal F_{\2}^2\big)\,,\cr
G_2&=&e^{-\varphi_1+\varphi_2+\varphi_3}{*F}_{\2,2}+\chi_1\mathcal F_{\2}^2\,,\cr
G_3&=&e^{-\varphi_1-\varphi_2+\varphi_3}{*\mathcal F}_{\2}^1
+\chi_2e^{-\varphi_1+\varphi_2+\varphi_3}*F_{\2,2}
+\chi_1\big(F_{\2,1}+\chi_2\mathcal F_{(2)}^2\big)\,,\cr
G_4&=&e^{-\varphi_1-\varphi_2-\varphi_3}{*\mathcal F}_{\2}^2
-\chi_2 e^{-\varphi_1+\varphi_2-\varphi_3}{*F}_{\2,1}
+\chi_3 e^{-\varphi_1-\varphi_2+\varphi_3}{*\mathcal F}_{\2}^1\cr
&&+\chi_2\chi_3\mathcal G_{\2}
+\chi_1\big(F_{(2),2}+\chi_3F_{\2,1}
-\chi_2\mathcal F_{\2}^1+\chi_2\chi_3\mathcal F_{\2}^2\big)\,.
\eea
It follows that the electric and magnetic charges are given by
\be
Q_{eI}=-\frac{1}{16\pi}\int G_I\,,\qquad
Q_{mI}=\frac{1}{16\pi}\int dA_I\,.
\ee
Note that the above is based on the Lagrangian \eqref{lag} from the $T^2$ reduction discussed in Section \ref{sec:theory}. The role of electric or magnetic charges can change if we use different ${\cal L}_{\rm K}$ Lagrangian. Thus, as mentioned above, the terms electric and magnetic are generally ambiguous. For our dyonic black holes, however, each gauge field carries unambiguously one electric and one magnetic charge.

To express the electromagnetic potentials, we take the values of the linear functions \eqref{8 charges linear} at \(r=r_+\) and \(u=0\), denoting them as
\bea
\mathcal V_{i+}
&=&q_i r_+-\frac{\mu n_1}{\mathcal N}\lambda_i+\mu\mathcal C_i\,,\qquad
\mathcal L_{i+}
=\frac{n_2\alpha_i+n_1\beta_i}{\mathcal N}r_+
  +2\mu\mathcal D_i+\frac{\mu n_1}{\mathcal N}p_i,\cr
\tilde{\mathcal V}_{i+}
&=&p_i r_+-\frac{\mu n_1(n_2\alpha_i+n_1\beta_i)}{\mathcal N^2}
  +\mu\tilde{\mathcal C}_i\,,\qquad
\tilde{\mathcal L}_{i+}
=-\lambda_i r_+-2\mu\tilde{\mathcal D}_i
  -\frac{\mu n_1}{\mathcal N}q_i\,.\label{8 charges linear horizon u0}
\eea
The electric and magnetic potentials conjugate to electric and magnetic charges are
\bea
\Phi_i&=&\frac{2\mu\sqrt{\Xi}}{F(r_+)}\Big[\big(R(r_+)+a^2\big)\mathcal{V}_{i+}+L(r_+)\mathcal{L}_{i+}\Big]\,,\cr
\Psi_i&=&\frac{2\mu\sqrt{\Xi}}{F(r_+)}\Big[\big(R(r_+)+a^2\big)\tilde{\mathcal{V}}_{i+}+L(r_+)\tilde{\mathcal{L}}_{i+}\Big]\,.\label{8 charges Phi Psi}
\eea
Having defined these notations, we list the electric and magnetic potentials together with the electric and magnetic dual charges of the four \(U(1)\) gauge fields in the following table
\bea
\renewcommand{\arraystretch}{1.4}
\setlength{\tabcolsep}{14pt}
\begin{tabular}{ccccc}
  \hline
  \rule{0pt}{24pt}
  & $\displaystyle\frac{2\sqrt{\Xi}}{\mu}Q_{eI}$
  & $\displaystyle\frac{2\sqrt{\Xi}}{\mu}Q_{mI}$
  & $\Phi_{eI}$ & $\Phi_{mI}$ \\[6pt]
  \hline
  $A_{\1,1}$ & $p_1$ & $q_1$ & $\Psi_1$ & $\Phi_1$ \\
  \hline
  $A_{\1,2}$ & $q_2$ & $-p_2$ & $\Phi_2$ & $-\Psi_2$ \\
  \hline
  $\mathcal{A}_{\1}^1$ & $p_3$ & $q_3$ & $\Psi_3$ & $\Phi_3$ \\
  \hline
  $\mathcal{A}_{\1}^2$ & $q_4$ & $-p_4$ & $\Phi_4$ & $-\Psi_4$ \\
  \hline
\end{tabular}\label{table}
\eea
Note that the electric and magnetic charges $(Q_{eI}, Q_{mI})$ are indeed directly proportional to $(q_i,p_i)$ given in \eqref{8pqs}; which is why we sometimes refer to $(q_i,p_i)$ simply as charges. As discussed above, to satisfy the first law, we also need to take into account the boundary contributions from the scalar fields \cite{Lu:2013ura,Chow:2013gba}. Gauged STU supergravity contains three dilatons and three axions, so we need to introduce six thermodynamic conjugate pairs. We introduce a constant parameter \(b=\frac{\mu}{\mathcal{N}}(n_2m_1-n_1m_2-n_2)\) and define the following coefficients
\bea
\sigma_A
&=&\frac{2\mu}{\mathcal N}(n_2\eta_{A1}+n_1\eta_{A2})-2b\,,\qquad
 \tau_A=\frac{2\mu}{\mathcal N}(n_2\xi_{A1}+n_1\xi_{A2})\,,\cr
\sigma_A^{\2}
&=&\mu^2\Big[
   \frac{n_1^2}{\mathcal N^2}(1+2\eta_{A1})
   -\frac{2n_1n_2}{\mathcal N^2}\eta_{A2}+4\eta_{A3}
  \Big]-b^2-b\sigma_A-\frac12(\sigma_A^2+\tau_A^2)
      +\frac14\sum_{B=1}^3(\sigma_B^2+\tau_B^2)\,,\cr
\tau_A^{\2}
&=&2\mu^2\Big[
   \frac{n_1^2}{\mathcal N^2}\xi_{A1}
   -\frac{n_1n_2}{\mathcal N^2}\xi_{A2}+2\xi_{A3}
  \Big]-b\tau_A\,.\label{8 charges scalar first law constants}
\eea
The contributions from the three dilatons and three axions are
\be
Y_{\varphi A}=\frac{\sigma_A^{\2}}{\sigma_A^2}\,,\qquad X_{\varphi A}=\frac{g^2}{12\Xi}\sigma_A^3\,,\qquad
Y_{\chi A}=\frac{\tau_A^{\2}}{\tau_A^2}\,,\qquad X_{\chi A}=\frac{g^2}{12\Xi}\tau_A^3\,.
\ee
It is now a straightforward but involved exercise to verify that the first law \eqref{firstlaw} augmented by \eqref{Zform} is precisely satisfied.

\subsection{Three-equal $U(1)$ supergravity}

Below, we present some thermodynamic quantities for the black hole solution in the three-equal truncation described by the Lagrangian \eqref{S3 action}
\bea
M&=&\frac{\mu}{\Xi\mathcal{N}}(n_2m_1-n_1m_2)\,,\quad S=\frac{\pi}{\Xi}\sqrt{F(r_+)}\,,\quad T=\frac{\Delta_r'(r_+)}{4\pi\sqrt{F(r_+)}}\,,\quad
J=\frac{\mu a}{\Xi}\mathcal{N}\,,\cr
\Omega&=&\frac{a\Xi L(r_+)}{F(r_+)}\,,\qquad \mathcal{P}=\frac{3g^2}{8\pi}\,,\qquad V=\frac{\pi F'(r_+)}{3\Xi}+\frac{4\pi a^2}{3}M\,.
\eea
It can be seen that, compared with the most general eight-charge case \eqref{8 charges M J}, the expressions for the mass and angular momentum simplify, while the remaining quantities retain the same form.

Similarly, we define the values of the linear functions appearing in the gauge fields at the horizon \(r=r_+\), \(u=0\) \eqref{8 charges linear horizon u0}. Although \(\mathcal{C}_{1,2},\mathcal{D}_{1,2}\) already suffice for the exact solution \eqref{S3 gauged linear}, an electromagnetic duality transformation was performed on \(A_{1,2}^{\mathrm{STU}}\) when the action \eqref{S3 action} was written down; consequently, the additional quantities \(\tilde{\mathcal{C}}_{1,2},\tilde{\mathcal{D}}_{1,2}\) are still needed in order to compute the complete electric and magnetic potentials
\bea
\tilde{\mathcal C}_1
&=&-2(n_2\alpha_1+n_1\beta_1)\cr
&&+\frac{1}{\mathcal N}\Big\{
  [n_2(4m_1-2-\eta_1)-n_1(4m_2+\eta_2)]q_1
 +(n_2\xi_1+n_1\xi_2)(2p_1-p_2)+n_1\lambda_1
 \Big\}\,,\cr
\tilde{\mathcal C}_2
&=&2\mathcal N\lambda_2+\beta_2+\frac{1}{\mathcal N}\Big\{
  [n_2(6m_1-3-3\eta_1)-n_1(6m_2+3\eta_2)]p_2
  -3(n_2\xi_1+n_1\xi_2)q_1
  \Big\},\cr
\tilde{\mathcal{D}}_1&=&s_1c_1^2c_2\hat{s}_1^3\hat{s}_2+c_1s_1^2s_2\hat{c}_1^3\hat{c}_2+s_1c_2(2c_1^2+s_1^2)\hat{s}_1^2\hat{c}_1\hat{c}_2
+c_1s_2(c_1^2+2s_1^2)\hat{c}_1^2\hat{s}_1\hat{s}_2\,,\cr
\tilde{\mathcal{D}}_2&=&s_2^2\hat{s}_2\hat{c}_1^3-c_2^2\hat{c}_2\hat{s}_1^3\,,
\eea
We next present the values at \(r=r_+\), \(u=0\) of the gauge-field linear functions \eqref{S3 gauged linear} together with those of their electromagnetic-dual counterparts
\bea
\mathcal V_{1+}
&=&p_1r_+-\frac{\mu n_1(n_2\alpha_1+n_1\beta_1)}{\mathcal N^2}
  +\mu\mathcal C_1\,,\qquad
\mathcal L_{1+}
=-\lambda_1r_+-2\mu\mathcal D_1-\frac{\mu n_1}{\mathcal N}q_1\,,\cr
\mathcal V_{2+}
&=&q_2r_+-\frac{\mu n_1}{\mathcal N}\lambda_2+\mu\mathcal C_2\,,\qquad
\mathcal L_{2+}
=\frac{n_2\alpha_2+n_1\beta_2}{\mathcal N}r_+
  +2\mu\mathcal D_2+\frac{\mu n_1}{\mathcal N}p_2\,,\cr
\tilde{\mathcal V}_{1+}
&=&q_1r_+-\frac{\mu n_1}{\mathcal N}\lambda_1
  +\mu\tilde{\mathcal C}_1\,,\qquad
\tilde{\mathcal L}_{1+}
=\frac{n_2\alpha_1+n_1\beta_1}{\mathcal N}r_+
  +2\mu\tilde{\mathcal D}_1+\frac{\mu n_1}{\mathcal N}p_1\,,\cr
\tilde{\mathcal V}_{2+}
&=&p_2r_+-\frac{\mu n_1(n_2\alpha_2+n_1\beta_2)}{\mathcal N^2}
  +\mu\tilde{\mathcal C}_2\,,\qquad
\tilde{\mathcal L}_{2+}
=-\lambda_2r_+-2\mu\tilde{\mathcal D}_2
  -\frac{\mu n_1}{\mathcal N}q_2\,.
\eea
Having introduced the above notation, we continue to employ the notation \eqref{8 charges Phi Psi} of the eight-charge case for the electric and magnetic potentials
\bea
Q_{e1}&=&\frac{\sqrt{3}\mu}{2\sqrt{\Xi}}p_1\,,\qquad Q_{m1}=\frac{\sqrt{3}\mu}{2\sqrt{\Xi}}q_1\,,\qquad
Q_{e2}=\frac{\mu}{2\sqrt{\Xi}}q_2\,,\qquad Q_{m2}=-\frac{\mu}{2\sqrt{\Xi}}p_2\,,\cr
\Phi_{e1}&=&\sqrt{3}\Phi_1\,,\qquad
\Phi_{m1}=\sqrt{3}\Psi_1\,,\qquad
\Phi_{e2}=\Phi_2\,,\quad
\Phi_{m2}=-\Psi_2\,.
\eea
Analogously to the previous case, we introduce a constant \(b=\frac{\mu}{\mathcal{N}}(n_2m_1-n_1m_2-n_2)\) and give the first-order and second-order asymptotic coefficients
\bea
\sigma
&=&2\sqrt{3}\Big[b-\frac{\mu}{\mathcal N}(n_2\eta_1+n_1\eta_2)\Big]\,,\qquad
 \tau=\frac{2\sqrt{3}\mu}{\mathcal N}(n_2\xi_1+n_1\xi_2)\,,\cr
\sigma^{(2)}
&=&\sqrt{3}\Bigg\{
 b^2-\mu^2\Big[
   \frac{n_1^2}{\mathcal N^2}(1+2\eta_1)
  -\frac{2n_1n_2}{\mathcal N^2}\eta_2+4\eta_3
 \Big]\Bigg\}
 -b\sigma-\frac{\sigma^2+\tau^2}{4\sqrt{3}}\,,\cr
\tau^{(2)}
&=&2\sqrt{3}\mu^2\Big[
 \frac{n_1^2}{\mathcal N^2}\xi_1
 -\frac{n_1n_2}{\mathcal N^2}\xi_2+2\xi_3
 \Big]-b\tau \,
\eea
while the contribution of the scalars to the first law is
\be
Y_\varphi=\frac{\sigma^{\2}}{\sigma^2}\,,\qquad X_\varphi=\frac{g^2}{12\Xi}\sigma^3\,,\qquad
Y_\chi=\frac{\tau^{\2}}{\tau^2}\,,\qquad X_\chi=\frac{g^2}{12\Xi}\tau^3\,.
\ee
Therefore, the first law can now be written as
\be
\delta M = T\delta S+\Omega\delta J+\sum_{i=1}^2(\Phi_{ei}\delta Q_{ei}+\Phi_{mi}\delta Q_{mi})+V\delta \mathcal{P}+X_\varphi\delta Y_\varphi+X_\chi\delta Y_\chi\,.
\ee

\subsection{$U(1)^2$ truncation}

Here, we study the thermodynamics of the black hole described by \eqref{Two charges truncation action}. In terms of the functions \eqref{Two charges truncation functions} and constants \eqref{Two charges truncation constants}, some of the thermodynamic quantities can be expressed as
\bea
M&=&\frac{\mu}{2\Xi}(d_1+d_2)\,,\qquad S=\frac{\pi}{\Xi}\sqrt{F(r_+)}\,,\qquad T=\frac{\Delta_r'(r_+)}{4\pi\sqrt{F(r_+)}}\,,\cr
J&=&\frac{\mu aD}{\Xi\eta}\,,\qquad \Omega=\frac{a\Xi L(r_+)}{F(r_+)}\,,\qquad \mathcal{P}=\frac{3g^2}{8\pi}\,,\qquad V=\frac{\pi F'(r_+)}{3\Xi}+\frac{4\pi a^2}{3}M\,.
\eea
Likewise, these results are consistent with the most general eight-charge case \eqref{8 charges M J}.

To express the electric and magnetic potentials, we follow \eqref{KK H and Phi} and introduce notation for the values taken by \(H_i(r,u)\) on the horizon \(r=r_+\) and on the equatorial plane \(u=0\)
\bea
H_{i+}=H_i(r_+,0)=r_+\big[r_++2\mu(d_{\bar{i}}-1)\big]+\mu^2(1-\eta^2)\,.
\eea
The charge and electric potential of the black hole are
\bea
Q_{ei}&=&\frac{\mu q_i}{2\sqrt{\Xi}}\,,\qquad Q_{mi}=-\frac{\mu p_i}{2\sqrt{\Xi}}\,,\cr
\Phi_{ei} &=& \frac{2\mu\sqrt{\Xi}}{F(r_+)}\Big[
(H_{i+}+a^2)\big(q_ir_++\frac{\mu p_i\mathcal{J}_i}{d_{\bar{i}}}\big)-\frac{2\mu a^2\mathcal{J}_i}{\eta^2}\big(p_i+\frac{q_i\mathcal{J}_i}{d_i}\big)
\Big]\,,\cr
\Phi_{mi} &=& -\frac{2\mu\sqrt{\Xi}}{F(r_+)}\Big[
(H_{\bar{i}+}+a^2)\big(p_ir_++\frac{\mu q_i\mathcal{J}_i}{d_i}\big)-\frac{2\mu a^2\mathcal{J}_i}{\eta^2}\big(q_i+\frac{p_i\mathcal{J}_i}{d_{\bar{i}}}\big)
\Big]\,.
\eea
The theory contains two dilatons and one axion, all three of which contribute to the first law at the boundary. In this case, the constant \(b\) and the expansion coefficients \eqref{8 charges scalar first law constants} reduce to
\bea
b&=&\frac{\mu}{2}(d_1+d_2-2)\,,\qquad \sigma_1=\mu\sum_{i=1}^2\Big(\frac{p_i^2}{d_{\bar{i}}}-\frac{q_i^2}{d_i}\Big)\,,\qquad \sigma_2=\sqrt{2}\mu(d_2-d_1)\,,\cr
\tau&=&2\mu\Big(\frac{q_1p_2}{d_1}+\frac{q_2p_1}{d_2}\Big)\,,\qquad \sigma_1^{(2)}=-b\sigma_1+\frac{2\mu^2\mathcal{I}}{D^2}(p_1p_2-q_1q_2)\,,\qquad
\sigma_2^{(2)}=-b\sigma_2\,,\cr
\tau^{(2)}&=&-b\tau+\frac{2\mu^2\mathcal{I}\mathcal{S}}{D^2}\,.
\eea
These parameters can be combined to give the contribution of the scalars to the first law
\bea
X_{\varphi A}&=&\frac{g^2}{12\Xi}\sigma_A^3\,,\qquad Y_{\varphi A}=\frac{\sigma_A^{\2}}{\sigma_A^2}\,,\qquad A=1,2\,,\cr
X_\chi&=&\frac{g^2}{12\Xi}\tau^3\,,\qquad Y_{\chi}=\frac{\tau^{\2}}{\tau^2}\,.
\eea
The first law is
\bea
\delta M=T\delta S+\Omega \delta J+\sum_{i=1}^2(\Phi_{ei}\delta Q_{ei}+\Phi_{mi}\delta Q_{mi})+V\delta \mathcal{P}+\sum_{A=1}^2X_{\varphi A}\delta Y_{\varphi A}+X_\chi\delta Y_\chi\,.
\eea

\subsection{AdS Kaluza-Klein case}

For the gauged KK theory, we have adopted the new parametrization \eqref{KK EMD sc beta}. At this point, the mass and angular momentum are
\be
M=\frac{\mu}{\Xi}\frac{(1-\beta_1)(1-\beta_2)(1-\beta_1\beta_2)}{\Gamma_1\Gamma_2}\,,\qquad
J=\frac{\mu a}{\Xi}\frac{1+\beta_1\beta_2}{\sqrt{\Gamma_1\Gamma_2}}\,.
\ee
The remaining geometric quantities are also consistent with the most general eight-charge case \eqref{8 charges M J}
\bea
T&=&\frac{\Delta_r'(r_+)}{4\pi\sqrt{F(r_+)}}\,,\qquad S=\frac{\pi}{\Xi}\sqrt{F(r_+)}\,,\qquad \Omega=\frac{a\Xi L(r_+)}{F(r_+)}\,,\cr
\mathcal{P}&=&\frac{3g^2}{8\pi}\,,\qquad V=\frac{\pi F'(r_+)}{3\Xi}+\frac{4\pi a^2}{3}M\,.
\eea
For the Maxwell sector, we can obtain the electric and magnetic dual charges together with the two potentials. For convenience, we denote the values of \(H_{1,2}\) on the horizon \(r=r_+\) and on the equator \(\theta=\pi/2\), and thereby define the potential parameters \(\Phi_i\)
\bea
H_{i+}&=&r_+^2+\frac{4\mu\beta_i}{\Gamma_i}\big[(1-\beta_{\bar{i}})r_++\mu\beta_{\bar{i}}\big]\,,\cr
\Phi_i&=&\frac{2\sqrt{2\Xi}\mu}{F(r_+)}\frac{\sqrt{\beta_{\bar{i}}\Gamma_i}}{\Gamma_{\bar{i}}}\Big[
\big(r_++\frac{2\mu\beta_i(1-\beta_{\bar{i}})}{\Gamma_i}\big)(H_{i+}+a^2)-\frac{4\mu a^2\beta_i(1-\beta_i\beta_{\bar{i}})}{\Gamma_i^2}
\Big]\label{KK H and Phi}
\eea
The electric and magnetic dual charges and the dual potentials can be expressed as
\be
Q_e=\frac{\mu}{\sqrt{2\Xi}}\frac{\sqrt{\beta_2\Gamma_1}}{\Gamma_2}\,,\quad Q_m=-\frac{\mu}{\sqrt{2\Xi}}\frac{\sqrt{\beta_1\Gamma_2}}{\Gamma_1}\,,\quad
\Phi_e=\Phi_1\,,\quad
\Phi_m=-\Phi_2\,.
\ee
Since the theory \eqref{gauged KK EMD} contains only one dilaton, only one thermodynamic conjugate pair of variables \(X,Y\) needs to be introduced here
\be
Y=\frac{\Gamma_2}{\Gamma_1}\sqrt{\frac{\beta_1}{\beta_2}}\,,\qquad X=\frac{4g^2\mu^3(\beta_1-\beta_2)\sqrt{\beta_1\beta_2^3}}{\Xi(1-\beta_1\beta_2)\Gamma_2^2}\,
\ee
and the extended first law is
\be
\delta M=T\delta S+\Omega\delta J+\Phi_e\delta Q_e+\Phi_m\delta Q_m+V\delta \mathcal{P}+X\delta Y\,.
\ee
Note that we omitted the discussion of the thermodynamics of the pairwise-equal case, as it was well understood in \cite{Chow:2014cca}.

\section{Euclidean on-shell action}
\label{sec:onshell}

Although the matter sector of the gauged STU supergravity is of the garden-variety type, the charged rotating AdS black holes are generally not. The ten-parameter solution suffers two shortcomings. One is that the solutions generally involve a conic singularity associated with the azimuthal angle $\phi$. The second is that the scalar hairs give nontrivial contributions to the first law. Although our first law can be verified using the Iyer-Wald formalism, we would like to examine it using the on-shell Euclidean action approach \cite{Gibbons:1976ue}. We shall therefore also compute the free energy of each black hole using the on-shell action method and compare the result with the thermodynamic description based on the first law. Here we adopt the counterterm method for the calculation. The total on-shell action consists of the Euclidean bulk action, the Gibbons-Hawking boundary term, and the counterterm
\bea
I_{\mathrm{tot}}=\frac{1}{16\pi}\int_{ \mathcal{M}} d^4x\sqrt{-g}\mathcal{L}_{\mathrm{bulk}}-\frac{1}{8\pi}\int_{\partial \mathcal{M}}d^3x\sqrt{-h}
\mathcal{L}_{\mathrm{GH}}-\frac{1}{8\pi}\int_{\partial \mathcal{M}}d^3x\sqrt{-h}
\mathcal{L}_{\mathrm{ct}}\,.
\eea
Here \(\mathcal{L}_{\mathrm{bulk}}\) denotes the action of gauged STU supergravity, or that of any theory obtained from it through a consistent truncation. Since all these theories have second-order equations of motion, the Gibbons-Hawking boundary term takes the same form in all cases, \(\mathcal{L}_{\mathrm{GH}}=K\). Following the approach of \cite{Lu:2013ura}, we find that the scalar fields also affect the specific expressions of the counterterms. Therefore, we need to specify the corresponding counterterms for each truncated theory.

{\bf General 8-charge solution:} Now we shall compute the on-shell action for the general eight-charge rotating AdS black hole. We shall adopt the bulk action associated with the $T^2$-reduction Lagrangian \eqref{lag}. The corresponding counterterm is
\bea
\mathcal L_{\rm ct}
&=&2g+\frac{\mathcal R}{2g}
-\frac16\sum_{A=1}^3\Big[
\big(\varphi_A+\frac12e^{\varphi_A}\chi_A^2\big)
 n^\mu\partial_\mu\varphi_A
+e^{\varphi_A}\chi_A\,n^\mu\partial_\mu\chi_A
\Big]\cr
&&+\frac g{12}\sum_{A=1}^3
\big(\varphi_A^2+e^{\varphi_A}\chi_A^2\big)\,.
\eea
Here we should note that, owing to the presence of the deficit angle, in addition to changing the integration range of \(\phi\) to \([0,2\pi]\), the conical defect on the axis will in principle also contribute to the curvature and hence change the on-shell action
\be
T I_{E,\rm cone} =\frac{r_++b}{8\pi}
\big(\delta_{\rm N}\kappa_{\rm N}
+\delta_{\rm S}\kappa_{\rm S}\big)=\frac14 (r_++b) (\kappa_{\rm N} + \kappa_{\rm S} -2)\,.
\ee
Since the physical interpretation of this contribution is not entirely clear, in this paper we consider only the regularized on-shell action. We find that the regularized on-shell action
\be
T I_{E,\mathrm{reg}} =M-TS-\Omega J-\sum_{I=1}^{4}\Phi_{eI}Q_{eI}
\ee
coincides with the conventional result. The boundary contributions from the dilaton and axion, \(X_{\varphi,\chi A}\) and \(Y_{\varphi,\chi A}\), do not enter the free energy expression directly.

{\bf Three-equal truncation:} The boundary holographic counterterm for the action \eqref{S3 action} is
\bea
\mathcal L_{\rm ct}
&=&2g+\frac{\mathcal R}{2g}
-\frac{1}{6}\Big[
\big(\varphi-\frac{1}{2\sqrt3}e^{-\frac{\varphi}{\sqrt3}}\chi^2\big)
 n^\mu\partial_\mu\varphi
+e^{-\frac{\varphi}{\sqrt3}}\chi\,n^\mu\partial_\mu\chi
\Big]\cr
&&+\frac{g}{12}\big(\varphi^2+e^{-\frac{\varphi}{\sqrt3}}\chi^2\big)\,.
\eea
We find that the free energy given by the regularized Euclidean on-shell action agrees with the thermodynamics
\be
T I_{E,\mathrm{reg}} =M-TS-\Omega J-\sum_{i=1}^2\Phi_{ei}Q_{ei}\,.
\ee

{\bf $U(1)^2$ truncation:} The boundary holographic counterterm associated the bulk action \eqref{Two charges truncation action} is
\bea
\mathcal L_{\rm ct}
&=&2g+\frac{\mathcal R}{2g}
-\frac16\Big(\big(\varphi_1+\frac12e^{\varphi_1}\chi^2\big)
 n^\mu\partial_\mu\varphi_1 +\varphi_2\,n^\mu\partial_\mu\varphi_2
+e^{\varphi_1}\chi\,n^\mu\partial_\mu\chi \Big)\cr
&&+\frac g{12}\big(\varphi_1^2+\varphi_2^2+e^{\varphi_1}\chi^2\big).
\eea
The regularized on-shell action computed in this way is consistent with the first law
\be
T I_{E,\mathrm{reg}}
=M-TS-\Omega J-\sum_{i=1}^2\Phi_{ei}Q_{ei}\,.
\ee

{\bf AdS KK trunction:} The boundary counterterm for the gauged KK theory \eqref{gauged KK EMD} was already given in \cite{Lu:2013ura}
\be
\mathcal{L}_{\mathrm{ct}}=2g+\frac{1}{2g}\mathcal{R}-\frac{1}{6}\varphi n^\mu\partial_\mu\varphi+\frac{g}{12}\varphi^2\,.
\ee
In \cite{Lu:2013ura}, a static black hole was considered, and hence there was no deficit angle. In the rotating case considered here, regularization is still required, and the resulting on-shell action agrees with the first law
\be
TI_{E,\mathrm{reg}}=M-TS-\Omega J-\Phi_eQ_e\,.
\ee

\section{Entropy product rules and BPS-type extremal limits}
\label{sec:product}

In this section, we shall also examine the entropy-product rules first proposed in \cite{Cvetic:2010mn}. The horizons of the black hole are determined by the zeros of the function \(\Delta_r\) \eqref{8 charges metric functions}. Since \(\Delta_r\) is quartic in \(r\), there are a total of four horizons that exist, and a set of thermodynamic relations can be defined on each of them. For any such black hole, the product of the entropies can be written as
\bea
\prod_{\alpha=1}^{4}S_\alpha
=\frac{4\pi^4}{g^4} ( J^2+\mathcal I_4 ) \,.\label{entropy product}
\eea
Here \(r_\alpha\) denote all four roots of \(\Delta_r(r_\alpha)=0\), counted with multiplicity. For real horizons the positive area is chosen, while for complex-conjugate roots, the conjugate square root is taken, so that the product is always non-negative. The quantity \(J\) is the angular momentum of the black hole, and \(\mathcal{I}_4\) is a constant determined by the black hole charges, whose explicit form depends on the specific black hole.

In this section, we also take the BPS-type limit of our charged rotating AdS black holes. The term ``BPS-type'', coined in \cite{Lu:2013ura}, has two aspects. First, for charged AdS black holes in gauged supergravity, the BPS limit is not the same as the extremal limit. For example, in the case of an RN AdS black hole, the extremal limit corresponds to a black hole with zero temperature, while BPS limit leads to a solution with naked singularity, but with the mass-charge relation $M=Q$. (The latter case is the extremal black hole when we turn off the cosmological constant.) The second aspect is that our general dyonic black holes do not admit a BPS limit. One reason is that a single gauge field carrying both electric and magnetic charges, it has a D0-D6 origin, which is a non-supersymmetric brane intersection, although turning on NS-NS string $B$ fields can lead to supersymmetry \cite{Witten:2000mf}. Nevertheless, we can still take a BPS-type limit, which can yield supersymmetric solutions for suitable charge configurations. To illustrate this, we consider the static dyonic AdS black hole. The mass-charge relation in the BPS-type limit is
\be
M= (Q_e^{\fft23} + Q_m^{\fft23})^{\fft32}\,,\label{BPS type M Qem}
\ee
which is the same as the asymptotically-flat Gibbons-Wiltshire black hole. However, the dyonic AdS solution is no longer a black hole, but a spacetime with naked singularity, significantly different from the extremal limit. When $Q_m=0$, the solution is supersymmetric, when embedded in STU supergravity. Concretely, the BPS-type limit is obtained by taking $(\mu, a)$ to zero simultaneously and at the same rate, while keeping the charges $(Q_{eI}, Q_{mI})$ finite and nonzero.

{\bf General 8-charge solution:} We now prove \eqref{entropy product} for the eight-charge case; the subsequent truncated results can all be regarded as results obtained by specializing the charges in this eight-charge case. We define the quadratic function \(\Delta=R+a^2\), rewrite \(\Delta_r=\Xi\Delta+g^2F\) in \eqref{8 charges metric functions} accordingly, and thereby rewrite the expression for the entropy \eqref{8 charges M J} as follows
\bea
S_\alpha^2=\frac{\pi^2}{\Xi^2}F(r_\alpha)
=-\frac{\pi^2}{g^2\Xi}\Delta(r_\alpha)\,.
\eea
Letting \(z_\pm\) be the two roots of \(\Delta\), \(\Delta(z_\pm)=0\), from \(\Delta_r(r)=g^2\prod_\alpha(r-r_\alpha)\) we obtain
\bea
\prod_{\alpha=1}^4\Delta(r_\alpha)
=\frac{\Delta_r(z_+)\Delta_r(z_-)}{g^4}
=F(z_+)F(z_-)\,.
\eea
At the same time, we can also use \(\Delta\) to rewrite the function \(F\) in \eqref{8 charges metric functions}, \(F=\Delta(K-R+a^2)+L^2\). Therefore we have
\bea
\prod_{\alpha=1}^{4}S_\alpha^2
=\Big[
\frac{\pi^4}{g^4\Xi^2}
L(z_+)L(z_-)
\Big]^2 \,.
\eea
For the eight-charge solution, using directly \(z_++z_-=2\mu n_2/\mathcal{N}\) and \(z_+z_-=a^2-\mu^2n_1^2/\mathcal{N}^2\), we obtain
\bea
L(z_+)L(z_-)
=4\mu^2a^2\mathcal N^2
+4\mu^4\big[4\mathcal D_0(n_2+\mathcal D_0)-n_1^2\big]\,.
\eea
The first term of the above expression can be replaced by the eight-charge angular momentum \eqref{8 charges M J}. For the second term, we need the identity
\bea
4\mathcal D_0(n_2+\mathcal D_0)-n_1^2
=
4\Big(\prod_{I=1}^{4}q_I+\prod_{I=1}^{4}p_I\Big)
+\Big(\sum_{I=1}^{4}q_Ip_I\Big)^2
-2\sum_{I=1}^{4}(q_Ip_I)^2 \,.
\eea
After replacing \(p_I\) and \(q_I\) by \(Q_{eI}\) and \(Q_{mI}\), we obtain the form of \eqref{entropy product}. For the eight-charge black hole, \(\mathcal{I}_4\) in \eqref{entropy product} is
\bea
&&\mathcal I_4^{\mathrm{STU}}=16\Big[
4Q_{m1}Q_{e2}Q_{m3}Q_{e4}
+4Q_{e1}Q_{m2}Q_{e3}Q_{m4}\cr
&&+\big(Q_{e1}Q_{m1}-Q_{e2}Q_{m2}
      +Q_{e3}Q_{m3}-Q_{e4}Q_{m4}\big)^2
-2\sum_{I=1}^{4}(Q_{eI}Q_{mI})^2
\Big]\,.
\eea

We now consider taking the BPS-type limit. We introduce a scaling parameter $\epsilon$ by defining
\be
\mu=\epsilon\mu_0\,,\qquad
\hat s_4=\epsilon^{-1}\,,\qquad
\hat c_4=\sqrt{1+\epsilon^{-2}}\,,\qquad
a=\epsilon a_0 \,,
\ee
and then take $\epsilon\rightarrow0^+$. The remaining four \(\delta_I\) and the three \(\gamma_{1,2,3}\) are kept finite. Taking \(a_0=0\) gives the static limit. In this limit, we find that the constants \(m_{1,2}\) and \(n_{1,2}\) become
\bea
m_1&=&\epsilon^{-2}m_{1*}+O(1)\,,\qquad
n_1=\epsilon^{-2}n_{1*}+O(1)\,,\cr
m_2&=&\epsilon^{-1}m_{2*}+O(\epsilon)\,,\qquad
n_2=\epsilon^{-1}n_{2*}+O(\epsilon)\,.
\eea
We use \(*\) to denote the leading terms in the limit. For the index list \(i=1,2,3\), we use an overbar to denote the remaining two indices, e.g.~\(\bar{1}=23\). We define \(b_{I=1\sim4}\)
\be
b_{i=1\sim3}=\hat s_i\hat c_{\bar{i}}-\hat c_i\hat s_{\bar{i}}\,,
\qquad
b_4=\hat c_{123}-\hat s_{123}\,.
\ee
In these notations, we find that leading terms \(m_{1,2*}\) and \(n_{1,2*}\) can be written as
\bea
m_{1*}&=&\frac12(1+s_1^2+s_2^2+s_3^2-s_4^2)\,,\qquad
m_{2*}=\sum_{I=1}^4s_Ic_I b_I\,,\cr
n_{1*}&=&c_4s_{123}-s_4c_{123}\,,\qquad
n_{2*}=(c_{1234}-s_{1234})b_4+
\sum_{i=1}^{3}(c_{i4}s_{jk}-s_{i4}c_{jk})b_i .
\eea
In general, \(n_{1*}\neq 0\), the explicit results for \(q_{I*}=\lim_{\epsilon\to0}\epsilon q_I\) and \(p_{I*}=\lim_{\epsilon\to0}\epsilon p_I\) are
\bea
q_{I*}&=&
\frac{n_{2*}(1-2\delta_{I4})s_Ic_I
-n_{1*}(c_I^2+s_I^2)b_I}{|n_{1*}|}\,,\qquad
p_{I*}=
\frac{n_{1*}\partial_{\delta_I}n_{2*}
-n_{2*}\partial_{\delta_I}n_{1*}}{|n_{1*}|}\,.
\eea
The partial derivative of \(p_{I*}\) acts only on the two finite polynomials given above. Thus
\bea
(\mathfrak q_I,\mathfrak p_I)
&=&(Q_{eI,*},Q_{mI,*}) =\frac{\mu_0}{2}
\begin{cases}
(p_{I*},q_{I*})\,,& I=1,3\,,\\
(q_{I*},-p_{I*})\,,& I=2,4\,,
\end{cases}\\[2pt]
m_*&=&M_* =\mu_0\,\frac{n_{2*}m_{1*}-n_{1*}m_{2*}}{|n_{1*}|}\,,
\qquad
j_*=J_*=\mu_0a_0|n_{1*}|\,.\label{8 charges BPS parameters}
\eea
We define the following combinations of charges
\bea
w_1&=&\mathfrak p_1+i\mathfrak q_1\,,\qquad
w_2=\mathfrak q_2-i\mathfrak p_2\,,\qquad
w_3=\mathfrak p_3+i\mathfrak q_3\,,\qquad
w_4=\mathfrak q_4-i\mathfrak p_4\,,\cr
Z_0&=&\sum_{I=1}^4w_I,\qquad Z_{j=1\sim3}=2(w_j+w_4)-Z_0 \,.\label{w1234 Z0123}
\eea
Let \(\rho_j=|Z_j|\), and define \(\Sigma_{1,2,3,4}\) by means of
\be
\prod_{j=0}^{3}(x+\rho_j^2)
=x^4+\Sigma_1x^3+\Sigma_2x^2+\Sigma_3x+\Sigma_4
\ee
Then we define
\be
H=\Sigma_1^2-4\Sigma_2\,,\qquad
K=\Sigma_1H+8\Sigma_3\,,\qquad
d=H+8\operatorname{Re}(Z_0Z_1Z_2Z_3)=-4\mu_0^4n_{1*}^2<0\,.
\ee
The eight-charge limit taken corresponds to \(d<0\). The mass-charge relation is a twelve-order polynomial equation of mass $m_*$, with eight charges form the coefficients:
\be
A^2B^2-4B^3-4A^3C+18ABC-27C^2=0 \,,\label{8 charges mass formular}
\ee
where
\bea
A &=& 8m_*^2(2m_*^2-\Sigma_1)-2H\,,\quad B=H^2-64\Sigma_4+8m_*^2\big[K+(\Sigma_1-4m_*^2)d\big]\,,\cr
C &=& 8m_*^2d(2m_*^2d-K)\,.
\eea

{\bf Three-equal truncation:} In this case, we shall not present the detail of derivations, but simply give the answer. The quantity \(\mathcal{I}_4\) in \eqref{entropy product} is
\bea
\mathcal I_4^{\rm 3\text{-}equal}&=&16\Big(
\frac{4}{3\sqrt3}\big(Q_{m1}^{3}Q_{e2}-Q_{e1}^{3}Q_{m2}\big)
+\frac13(Q_{e1}Q_{m1})^2\cr
&&-2Q_{e1}Q_{m1}Q_{e2}Q_{m2}
-(Q_{e2}Q_{m2})^2
\Big)\,.
\eea
Setting \(Q_{e1}=Q_{m1}=0\) yields the gauged KK result discussed later.

For the BPS-type limit, the result can be obtained directly by taking
\bea
&&(\mathfrak q_1^{\mathrm{STU}},\mathfrak p_1^{\mathrm{STU}})
=(\mathfrak q_3^{\mathrm{STU}},\mathfrak p_3^{\mathrm{STU}})
=\frac{1}{\sqrt3}
  (\mathfrak q_1,
   \mathfrak p_1)\,,\cr
&&(\mathfrak q_2^{\mathrm{STU}},\mathfrak p_2^{\mathrm{STU}})
=\frac{1}{\sqrt3}
  (\mathfrak p_1,
  -\mathfrak q_1)\,,\qquad
(\mathfrak q_4^{\mathrm{STU}},\mathfrak p_4^{\mathrm{STU}})
=(\mathfrak q_2,
   \mathfrak p_2)\,.\label{three equal charges BPS parameters}
\eea
in \eqref{8 charges BPS parameters}. Afterwards we can define \(w\) and \(z\)
\be
w=\frac{\mathfrak p_1
        +i\mathfrak q_1}{\sqrt3}\,,\qquad z=\mathfrak q_2 -i\mathfrak p_2\,,
\ee
then \(Z_{0,1,2,3}\) in \eqref{w1234 Z0123} becomes
\be
Z_0=z+3w\,,\qquad Z_1=Z_2=Z_3=z-w\,.
\ee
Defining\(U=|z+3w|^2\), \(V=|z-w|^2\), and \(\mathcal{R}_3=\operatorname{Re}[(z+3w)(z-w)^3]\), we have
\bea
\Sigma_1&=&U+3V\,,\qquad \Sigma_2=3V(U+V)\,,\qquad
\Sigma_3=V^2(3U+V)\,,\cr
 \Sigma_4&=&UV^3\,,\qquad
d=U^2-6UV-3V^2+8\mathcal R_3.
\eea
By now, the mass formula \eqref{8 charges mass formular} is reduced to a six-order polynomial equation
\bea
64m_*^6-48(U+3V)m_*^4
+12(U^2-3UV+18\mathcal R_3)m_*^2
-U(U-9V)^2=0\,.\label{three equal charges mass formular}
\eea

For the $U(1)^2$ case, \(\mathcal{I}_4\) in \eqref{entropy product} is
\bea
\mathcal{I}_4^{\mathrm{two-U(1)}}=-16\big(Q_{e1}Q_{m1}-Q_{e2}Q_{m2}\big)^2
\eea

{\bf $U(1)^2$ truncation:} For this case, \(\mathcal{I}_4\) in \eqref{entropy product} becomes a rather simple expression
\be
\mathcal{I}_4^{\mathrm{2-U(1)}}=-16\big(Q_{e1}Q_{m1}-Q_{e2}Q_{m2}\big)^2\,.
\ee
The BPS-type limit, on the other hand, is still somewhat nontrivial. We first make a charge truncation  \eqref{8 charges BPS parameters} by taking
\be
\mathfrak q^{\mathrm{STU}}
=(0,\mathfrak q_2,0,\mathfrak q_1)\,,\qquad
\mathfrak p^{\mathrm{STU}}
=(0,\mathfrak p_2,0,\mathfrak p_1)\,.
\ee
Since \(w_1=w_3=0\). we have
\be
Z_0=Z_2=w_4+w_2\,,\qquad Z_1=Z_3=w_4-w_2\,.
\ee
We define the following three combinations
\be
\mathfrak s=\sum_{i=1}^2(\mathfrak q_i^2+
                     \mathfrak p_i^2)\,,\qquad
\mathfrak c=\mathfrak q_1\mathfrak q_2
  +\mathfrak p_1\mathfrak p_2\,,\qquad
\mathfrak J=\mathfrak q_1\mathfrak p_1
          -\mathfrak q_2\mathfrak p_2\,.
\ee
Consequently, the four \(\rho_j^2\) are pairwise equal
\be
\rho_0^2=\rho_2^2=\mathfrak s+2\mathfrak c\,,\qquad
\rho_1^2=\rho_3^2=\mathfrak s-2\mathfrak c\,,
\ee
and hence
\bea
\Sigma_1=4\mathfrak s\,,\qquad H=-8(\mathfrak s^2-4\mathfrak c^2)\,,\qquad
\Sigma_4=(\mathfrak s^2-4\mathfrak c^2)^2,\qquad K=0,\qquad d=-64\mathfrak J^2.
\eea
The physical mass is given by the following quartic expression
\bea
\mathfrak c^2(x^2-4\mathfrak c^2)^2
+m_*^2\mathfrak J^2 x(x^2-36\mathfrak c^2)
-27m_*^4\mathfrak J^4=0\,,\qquad x=m_*^2-\mathfrak s \,.
\eea

{\bf AdS KK truncation:} This is the simplest nontrivial truncation. The quantity \(\mathcal{I}_4\) in \eqref{entropy product} can be read off directly from the three-equal truncation, giving
\be
\mathcal{I}_4^{\mathrm{KK}}=-16Q_e^2Q_m^2\,.
\ee
The BPS-type limit can also be obtained directly from the three-equal case \eqref{three equal charges BPS parameters} by taking
\be
\mathfrak q_1^{\mathrm{three\text{-}equal}}=\mathfrak p_1^{\mathrm{three\text{-}equal}}=0\,,\qquad
\mathfrak q_2^{\mathrm{three\text{-}equal}}=\mathfrak q\,,\qquad
\mathfrak p_2^{\mathrm{three\text{-}equal}}=\mathfrak p\,.
\ee
At this point the four \(Z\)'s are completely identical
\be
Z_0=Z_1=Z_2=Z_3=\mathfrak q-i\mathfrak p\,.
\ee
Defining \(U=\mathfrak{q}^2+\mathfrak{p}^2\), we then have
\be
(\Sigma_1,\Sigma_2,\Sigma_3,\Sigma_4)
=(4U,6U^2,4U^3,U^4)\,,\quad
H=-8U^2\,,\quad K=0\,,\quad d=-64\mathfrak q^2\mathfrak p^2\,.
\ee
Substituting into \eqref{three equal charges mass formular} yields
\be
m_*
=\Big(|\mathfrak q|^{2/3}
       +|\mathfrak p|^{2/3}\Big)^{3/2}\,.
\ee
As can be seen, this result is exactly \eqref{BPS type M Qem}, and thus agrees with the result of \cite{Lu:2013ura}.

\section{Conclusion}
\label{sec:con}

In this paper, we constructed the most general family of exact solutions describing non-extremal rotating AdS black holes in four-dimensional gauged STU supergravity, carrying four independent electric charges and four independent magnetic charges. This work thus completes a decades-long program of constructing such solutions through successive studies of various special cases.

The gauged STU supergravity admits a variety of consistent truncations, including both supersymmetric and non-supersymmetric ones, and many special solutions are known in the literature. We systematically surveyed these truncations and constructed the corresponding solutions. While purely electric solutions and dyonic solutions in the pairwise truncations have been extensively studied, the dyonic solutions in several other truncations presented here are new.

A characteristic feature of the general eight-charge dyonic solution is that, whenever the constant \(v\) defined in \eqref{constant v} is nonzero, there are necessarily deficit angles at the north pole, the south pole, or both, leading to a conical singularity associated with the azimuthal angle. It therefore follows that a rotating AdS black hole free of conical singularities can carry at most seven independent charges. The residual \(SO(2)^3\) global symmetry of the gauged theory enables us to generate the general eight-charge solution from a minimal seed carrying five independent charges. However, the requirement of eliminating the conical singularity implies that the minimal seed for generating a regular rotating AdS black hole can carry at most four independent charges.

We analysed various properties of our general solution and its various truncations. We derived the thermodynamic quantities and obtained the first law. As already found for the static dyonic cases, there must be scalar contributions to the first law. We also find that the first law can be established even in the presence of conical singularities by simply taking the azimuthal angle to have 2$\pi$ period. We also verify the free energy by explicitly computing the Euclidean on-shell actions. We have also derived the entropy product rules and shown that the entropy products depend only on the angular momentum and charges, and are therefore quantized when these quantities are quantized.

Although we considered the BPS-type limit, we have not investigated genuine BPS (supersymmetric) limits of our solutions or compared them with the known BPS solutions in the literature, e.g.~\cite{Cacciatori:2009iz, Benini:2015eyy,Hristov:2018spe,Hristov:2019mqp}. In our BPS-type limit, the mass depends only on the charges, but not on the angular momentum. However, in the BPS solutions, the mass also depends on the angular momentum. This implies that there are other BPS-type limits that are worth further exploring.
It is therefore of great interest to investigate all possible BPS and BPS-type limits of our general eight-charge solution. Our general solution provides a non-extremal generalization of the known BPS solutions, thereby allowing studies of the AdS/CFT correspondence along the lines of \cite{Hristov:2018spe, Choi:2018fdc,Hristov:2019mqp, Cassani:2019mms,Choi:2019zpz,Nian:2019pxj} to be extended to a broader framework. It is well established that maximal gauged supergravity in four dimensions can be obtained through an $S^7$ reduction of eleven-dimensional supergravity \cite{deWit:1986oxb,Godazgar:2015qia}, although the reduction ansatz can be rather complicated. Gauged STU supergravity is a much simpler subsector \cite{Azizi:2016noi}; and it is therefore of great interest to investigate the eleven-dimensional origin of our general eight-charge solution. In particular, it would be interesting to determine whether the conical singularities persist or are resolved upon uplift to eleven dimensions.

\section*{Acknowledgement}

L.M.~is supported in part by National Natural Science Foundation of China (NSFC) grant No.~12447138, Postdoctoral Fellowship Program of CPSF Grant No.~GZC20241211, the China Postdoctoral Science Foundation under Grant No.~2024M762338, the National Key Research and Development Program No.~2022YFE0134300, and also by the Tianjin University Self-Innovation Fund New Inspiration Fund No.~2026XQM09934. H.L.~is supported in part by the NSFC grants No.~12375052 and No.~11935009. The work is also supported in part by the Tianjin University Self-Innovation Fund Extreme Basic Research Project Grant No.~2025XJ21-0007. We also made an extensive use of AI programs in both research and English proofreading.

\end{document}